\documentclass[a4paper, twocolumn, superscriptaddress, showkeys, floatfix,11pt, accepted=2025-01-16]{quantumarticle}
\pdfoutput=1
\usepackage{silence}
\usepackage[utf8]{inputenc}
\usepackage[english]{babel}
\usepackage[T1]{fontenc}
\usepackage{array}
\usepackage{multirow}
\usepackage{dcolumn}
\usepackage{hyperref}
\usepackage[pdftex]{graphicx}
\usepackage[normalem]{ulem}
\usepackage{url}
\usepackage{braket}
\usepackage{here}
\usepackage{amsmath}
\usepackage{amsfonts}
\usepackage{euscript}
\usepackage{mathrsfs}
\usepackage{amssymb}
\usepackage{amsthm}
\usepackage{mathcomp}
\usepackage{scalefnt}
\usepackage{dcolumn}
\usepackage{float}
\usepackage[ruled,linesnumbered,noend]{algorithm2e}
\usepackage[FIGTOPCAP,hang,nooneline]{subfigure}
\usepackage{cite}
\usepackage{tikz}
\usetikzlibrary{decorations.pathreplacing}
\usetikzlibrary{trees,quotes,angles}

\newcommand{\ctext}[1]{\raise0.2ex\hbox{\textcircled{\scriptsize{#1}}}}

\definecolor{Apricot}{HTML}{E39882}
\definecolor{GreenYellow}{HTML}{E9E492}
\definecolor{Lavender}{HTML}{B288B6}
\definecolor{LimeGreen}{HTML}{C5D696}
\definecolor{Orchid}{HTML}{6CB6AD}
\definecolor{CornflowerBlue}{HTML}{6D8FB9}
\definecolor{Maroon}{HTML}{E6C0D6}
\definecolor{Tan}{HTML}{DF759B}
\definecolor{Grayred}{HTML}{E26464}
\definecolor{Grayyellow}{HTML}{F2D649}

\hypersetup{
    unicode=false,          
    pdftoolbar=true,        
    pdfmenubar=true,        
    pdffitwindow=false,     
    pdfstartview={FitH},    
    pdfkeywords={keyword1} {key2} {key3}, 
    pdfnewwindow=true,      
    colorlinks=true,       
}
\SetKwInput{KwData}{\textbf{Input}}
\SetKwInput{KwResult}{\textbf{Output}}
\newcommand{\red}[1]{\textcolor{red}{#1}}

\newcommand\erase{\bgroup\markoverwith{\textcolor{red}{\rule[.5ex]{2pt}{0.4pt}}}\ULon}
\begin{document}
\title{Multiplexed Quantum Communication with Surface and Hypergraph Product Codes}
\author{Shin Nishio}
\email{parton@nii.ac.jp}
\affiliation{SOKENDAI (The Graduate University for Advanced Studies), 2-1-2 Hitotsubashi, Chiyoda-ku, Tokyo, 101-8430, Japan}
\affiliation{Okinawa Institute of Science and Technology Graduate University, Onna-son, Kunigami-gun, Okinawa, 904-0495, Japan}
\affiliation{National Institute of Informatics, 2-1-2 Hitotsubashi, Chiyoda-ku, Tokyo, 101-8430, Japan}
\orcid{0000-0003-2659-5930}
\author{Nicholas Connolly}
\affiliation{Okinawa Institute of Science and Technology Graduate University, Onna-son, Kunigami-gun, Okinawa, 904-0495, Japan}
\email{nicholas.connolly@oist.jp}
\author{Nicolò Lo Piparo}  
\affiliation{Okinawa Institute of Science and Technology Graduate University, Onna-son, Kunigami-gun, Okinawa, 904-0495, Japan}
\author{William John Munro}
\affiliation{Okinawa Institute of Science and Technology Graduate University, Onna-son, Kunigami-gun, Okinawa, 904-0495, Japan}
\affiliation{National Institute of Informatics, 2-1-2 Hitotsubashi, Chiyoda-ku, Tokyo, 101-8430, Japan} 
\author{Thomas Rowan Scruby}  
\email{thomas.scruby@oist.jp}
\affiliation{Okinawa Institute of Science and Technology Graduate University, Onna-son, Kunigami-gun, Okinawa, 904-0495, Japan}
\author{Kae Nemoto} 
\email{kae.nemoto@oist.jp}
\affiliation{Okinawa Institute of Science and Technology Graduate University, Onna-son, Kunigami-gun, Okinawa, 904-0495, Japan}
\affiliation{National Institute of Informatics, 2-1-2 Hitotsubashi, Chiyoda-ku, Tokyo, 101-8430, Japan} 
\maketitle
\begin{abstract}
Connecting multiple processors via quantum interconnect technologies could help overcome scalability issues in single-processor quantum computers. Transmission via these interconnects can be performed more efficiently using quantum multiplexing, where information is encoded in high-dimensional photonic degrees of freedom. We explore the effects of multiplexing on logical error rates in surface codes and hypergraph product codes. We show that, although multiplexing makes loss errors more damaging, assigning qubits to photons in an intelligent manner can minimize these effects, and the ability to encode higher-distance codes in a smaller number of photons can result in overall lower logical error rates. This multiplexing technique can also be adapted to quantum communication and multimode quantum memory with high-dimensional qudit systems.
\end{abstract}
\section{Introduction}
Quantum computers are expected to solve problems that are intractable using classical computation~\cite{365700, grover1996fast}, but these powerful quantum algorithms require high qubit counts and deep circuits to solve problems of interesting size~\cite{haner2016factoring, yoshioka2024hunting}. While the qubit counts of quantum processors have been increasing rapidly in recent years, various physical constraints impose limits on the possible size of a single quantum processor~\cite{krinner2019engineering, tamate2022toward}. \textit{Quantum interconnects} provide a resolution to this problem by allowing for the networking and cooperative operation of multiple quantum processors~\cite{awschalom2021development}, as well as the use of separate quantum memories~\cite{monroe2014large}, quantum repeaters~\cite{azuma2023quantum,munro2008high,jiang2009quantum,munro2012quantum} and networks~\cite{van2014quantum, muralidharan2016optimal, munro2022designing} in analogy with classical computing architectures. Optical systems are considered leading candidates for practical implementation of quantum interconnects~\cite{wang2016chip} and also as quantum memories~\cite{lvovsky2009optical} due to their long coherence times~\cite{cho2016highly}.

Due to the high noise levels inherent in quantum systems, large-scale quantum algorithms cannot be executed reliably without the use of quantum error correcting codes (QECCs)~\cite{calderbank1996good,gottesman1997stabilizer}, which enable fault-tolerant quantum computation (FTQC)~\cite{gottesman1998fault,kitaev2003fault}. Similarly, optical interconnects~\cite{awschalom2021development, wang2016chip} can suffer from high photon loss rates and so QECCs should be used to protect information transmitted through these channels. In principle, it may be preferable to use different codes for these different settings~\cite{nagayama2016interoperability}, but in practice, the transfer of information between these different codes may be challenging enough that it is easier to use only a single code. For instance, fault-tolerant logic with surface codes~\cite{bravyi1998quantum} has been very well studied~\cite{fowler2012surface, horsman2012surface}, while quantum Reed-Solomon codes~\cite{grassl1999quantum} provide efficient and loss-tolerant protection for transmission through optical channels, but it is not clear how to interface or switch between these two families of codes. Therefore, for performing distributed computation with interconnects, using surface codes (or their generalizations) for both computation and transmission~\cite{fowler2010surface} is a natural alternative to using multiple codes. This is much less efficient in the sense of error-correction capability for the communication part, but these overheads can be reduced using \textit{quantum multiplexing}~\cite{piparo2019quantum}.

\begin{figure*}[htbp]
    \begin{minipage}[htbp]{0.65\linewidth}
    \begin{center}
        \includegraphics[width=0.9\columnwidth]{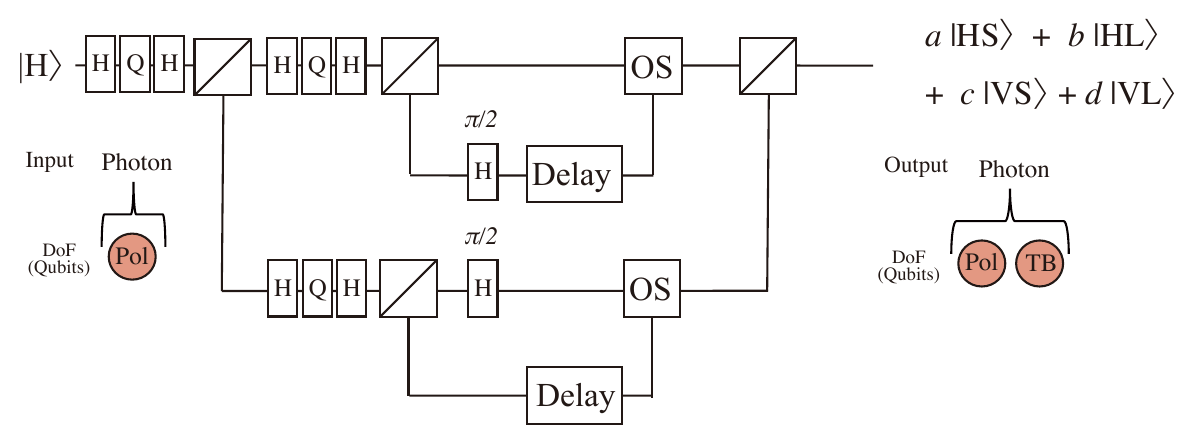}\\
        (A)\\
        \includegraphics[width=0.9\columnwidth]{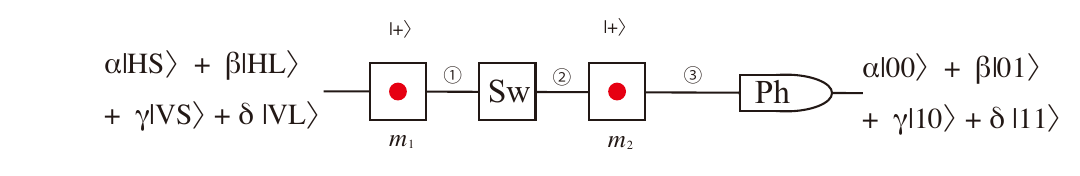}\\
        (B)\\
        \includegraphics[width=0.9\columnwidth]{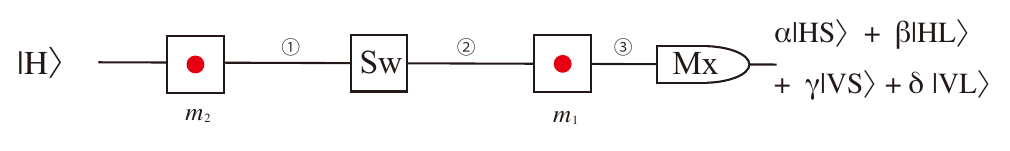}\\
        (C)\\
        \includegraphics[width=0.6\columnwidth]{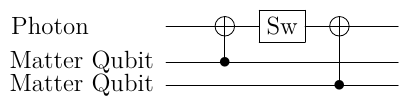}\\
        (D)\\
    \end{center}
    \end{minipage}
    \begin{minipage}[htbp]{0.9\linewidth}
        \includegraphics[width=0.4\columnwidth]{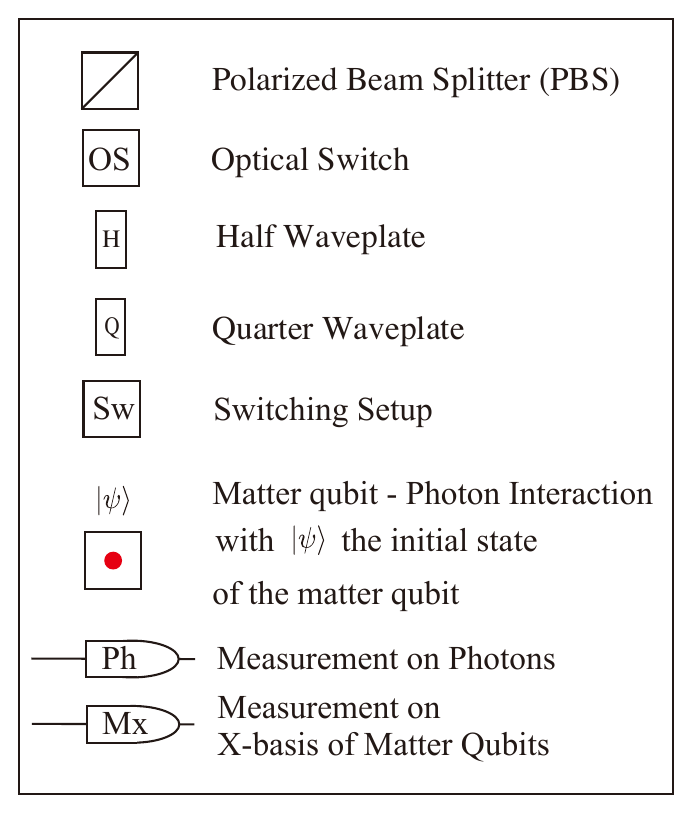}
    \end{minipage}
    \caption{ An example of an optical circuit encoding $2^2$-dimensional quantum information into a single photon. A horizontal (H) polarized photon is sent through a half-quarter-half wave plate appropriately rotated to generate a general polarized qubit. Then a polarizing beam splitter (PBS) divides the components of such a qubit into two spatial modes. Specifically, the horizontal component is transmitted whereas the vertical component is reflected. A delay is added to these components, and two optical switches and another PBS will recombine all the components into a single spatial mode.
    (B) A heralding method to transfer a general state of a multiplexed photon into two matter qubits $m_1$ and $m_2$ (red dots into boxes) by using a switching setup (Sw) module, which has been used in \cite{piparo2019quantum}. 
    (C) A heralding method to transfer a general state of two atoms into a multiplexed photon. Here a polarized photon interacts with an atom $m_1$ and then goes through a Sw module to add time-bin modes. Then this photon interacts with another atom $m_2$. We use $\alpha\ket{00} + \beta\ket{01}+\gamma\ket{10}+\delta\ket{11}$ for the input state of two-qubit memory. Upon a successful measurement in the $X$-basis of the two atoms, their state will be transferred into the multiplexed photon. 
    (D) The equivalent circuit of the setups in (B) ((C)), respectively. The interaction of a polarized photon with matter qubit $m_1$ ($m_2$) and then with matter qubit $m_2$ ($m_1$) corresponds to a CNOT gate in which the atom is the control qubit and the polarization DOF of the photon is the target qubit. Although not explicitly shown, at the end of the circuit the photon and matter qubits are measured for the setup in (B) and (C), respectively.}
    \label{fig:qm_encoder}
\end{figure*}

Quantum multiplexing is a technique for encoding high-dimensional quantum information into a single photon by exploiting multiple different photonic degrees of freedom (DOF) or a single multi-component degree of freedom. Such encodings can be performed using only linear optical elements and can significantly reduce the resources associated with quantum communication~\cite{nishio2023resource,piparo2020resource, piparo2020aggregating}. In this work, we examine the potential of quantum multiplexing for enabling efficient transmission of surface and hypergraph product (HGP) codes through optical channels. Densely encoding many of these codes' qubits into small numbers of photons has the potential to make loss errors much more damaging, but we present various techniques (e.g. optimized strategies for qubit-to-photon assignment) that can mostly or completely eliminate these downsides. 

The rest of this paper is organized as follows. In Sec.~\ref{sec_backkgrounds} we review the relevant background for quantum multiplexing and communication over lossy optical channels. Then in Sec.~\ref{sec_mqcwithecc} we propose three approaches to error-corrected quantum communication using multiplexing that provide different ways of reducing the impact of loss errors. This is followed up in Sec.~\ref{sec_SC} and Sec.~\ref{sec_HGP} with an examination of some of these approaches in more detail for surface codes and HGP codes, respectively. Finally, we discuss our findings and conclude this work in Sec.~\ref{sec_conc}. 

\section{Background}
\label{sec_backkgrounds}
In this section, we provide a brief overview of quantum multiplexing and erasure correction and show how these elements appear in practical quantum communication protocols.

\subsection{Quantum Multiplexing}
\label{sec_QM}
This subsection outlines the concept of quantum multiplexing and illustrates its possible implementation with an example.

In photon-based quantum information processing, various DOFs can be utilized to encode qubits. Polarizations~\cite{u2003photon}, time-bins~\cite{brendel1999pulsed, marcikic2002femtosecond, thew2002experimental}, paths (dual rail)~\cite{kok2007linear}, orbital angular momentum~\cite{yao2011orbital}, and frequency-bin~\cite{shih1994observation, ramelow2009discrete} are typical examples of DOFs in a single photon which are commonly used in experiments. Multi-level time-bins make it especially easy to encode high-dimensional quantum information in a single photon. For instance, Fig.~\ref{fig:qm_encoder}~(A) shows a method for encoding higher dimensional information (dimension $2^2$ for two qubits) using polarization and time-bin DOF. The series of Half-Quarter-Half waveplates as shown in Fig.~\ref{fig:qm_encoder}~(A) as HQH allows one to apply an arbitrary unitary gate to the polarization DOF of a photon. This corresponds to applying rotation gates around a Z-X-Z basis in a qubit. We label the horizontal and vertical components of a polarized photon with $H$ and $V$, respectively, and the short and long components of a temporal mode (time-bin) with $S$ and $L$, respectively.

This circuit takes a photon whose polarization is encoded with quantum information as input. This input photon has one qubit of information. After passing through this circuit, the photonic state $a\ket{HS}+b\ket{HL}+c\ket{VS}+d\ket{VL}$ has both polarization and time-bin degrees of freedom.

The polarization encodes a 2-dimensional Hilbert space, and the time-bin encodes another 2-dimensional Hilbert space. Therefore, the Hilbert space of the final encoded state has dimension $4$, thus encoding $2$ qubits of information. This encoding can easily be generalized to higher dimensional multiplexed photons as shown in Appendix~\ref{app:qmencoder}.
Significantly, encoding high-level time-bin states only requires linear optical elements and classical optical switches.

Quantum multiplexing~\cite{piparo2019quantum} is a method to encode higher dimensional quantum information into a single photon using these multiple degrees of freedom. 
One can transfer the state of this multiplexed photon embedded with an extra DOF (time-bin) into two matter qubits \cite{piparo2019quantum} as shown in Fig.~\ref{fig:qm_encoder}~(B). First, the photon is entangled with the internal DOFs of a matter qubit, depicted as a red dot in a box in the Figure. Then the $HL$ and $VS$ components of the photon are switched by the switching setup module (Sw). This is a slightly modified setup in \cite{piparo2019quantum} where we use conventional PBSs rather than 45\textdegree-PBSs. Next, the photon interacts with another matter qubit. Here we label the basis states of a matter qubit $\ket{0}$ and $\ket{1}$. After measuring the DOFs of the photon, the state of the photon is projected into the matter qubits upon a Pauli frame correction depending on the measurement outcome (write-in). This procedure can be extended to photons with an arbitrary number of time-bin components by adding the corresponding number of matter qubits. 
We can recover the information stored in the matter qubits with the process shown in Fig.~\ref{fig:qm_encoder}~(C) (read-out).
This process can be regarded as the inverse of Fig.~\ref{fig:qm_encoder}~(B). Here, a polarized photon interacts with a matter qubit $m_2$. Next, the photon goes through the Sw module which switches the $HL$ component and the $VS$ component. Then, the photon interacts with another matter qubit $m_1$. Finally, both matter qubits are measured in the $X$-basis. This projects the state of the matter qubits into the multiplexed photon. We give a quantum circuit representation in Fig.~\ref{fig:qm_encoder}~(D), and the full description of the state transition in the Appendix \ref{app:qmencoder2} for Fig.~\ref{fig:qm_encoder}~(B) and (C).

In this work, we consider encoding $2^m$-dimensional quantum information using $m$ components of a DOF per photon where $m$ is an integer ($m=1$ corresponds to no multiplexing). 

It is worth noticing that while quantum multiplexing allows for efficient communication, it also changes the error model. In fact, in a lossy communication channel, the loss of a photon causes the simultaneous loss of multiple qubits encoded in that photon. This can be very detrimental to the performance of this system. However, in the next section, we will devise several strategies for qubit assignment to mitigate the effects of the loss of qubits.

\subsection{Erasure Channel and Correction}
\label{sec_ER}
Let us now describe the erasure channel and decoding, which will play an important role in the quantum communication protocol.

In photonic systems, the erasure error is a localized loss error of a photon due to imperfections in the photon source, the physical channel used for its transmission, and detectors. This is the dominant source of errors in optical systems~\cite{slussarenko2019photonic, joshi2021quantum}.
Moreover, theoretical~\cite{wu2022erasure, kubica2023erasure, kang2023quantum, tsunoda2023error} and experimental~\cite{lu2008experimental, ma2023high, scholl2023erasure, levine2023demonstrating, chou2023demonstrating} works have been proposed on methods to map errors from different sources to erasure errors in multiple physical systems recently. Therefore, correction of erasure errors is of engineering importance because it can be applied in a variety of systems where erasure errors are not the main source of error.

The erasure channel is given by 
\begin{equation}
    \rho \rightarrow (1-\varepsilon) \rho + \varepsilon \ket{e}\bra{e}
\end{equation}
where $\ket{e}$ indicates the erased state, and $\varepsilon$ is the probability of erasure. Due to the fact that the erased state is not in the original Hilbert space, it is possible to detect such errors without further damaging the encoded quantum information. 

Several methods have been proposed to detect and correct erasure errors with QECCs~\cite{alber2001stabilizing}. It is possible to correct erasure by deforming the original logical operator~\cite{stace2009thresholds, barrett2010fault}, as well as by converting erasure errors into random Pauli errors by replacing the lost qubits with mixed states:
\begin{equation}
    \frac{\mathbb{I}}{2}=  \frac{1}{4}(\rho+X \rho X+Y \rho Y+Z \rho Z).
\end{equation}

After replacing the qubits, one can perform stabilizer measurements as normally occurs in surface codes. Then, the erasure is converted into random Pauli errors with the exact probabilities ($1/4$) for $\{I, X, Y, Z\}$. This random Pauli can also be regarded as independent $X$ and $Z$ errors with a probability of $1/2$.
This allows for the decoding of an erasure error.
The (surface code) peeling decoder~\cite{delfosse2020linear}, which is a linear-complexity erasure decoder using this procedure, has been proposed as a maximum-likelihood decoder for erasure errors in the surface code. We briefly overview the peeling decoder and its surface code generalization in Appendix~\ref{app:peeling}.

\subsection{Applying quantum multiplexing to an error-corrected erasure channel}
\begin{figure*}[htb]
    \centering
    \includegraphics[width=1.9\columnwidth]{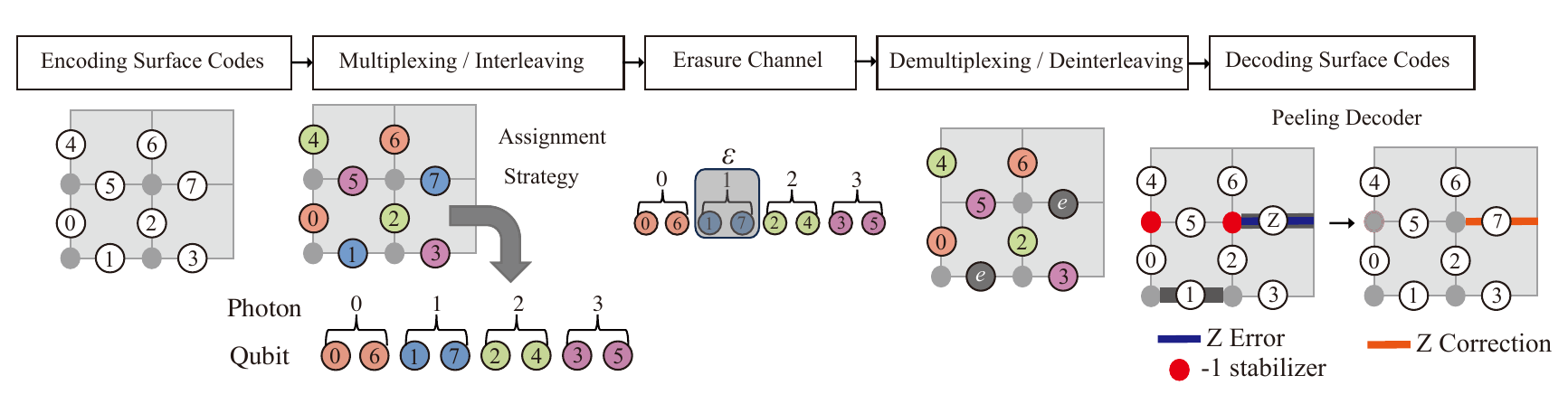}
    \caption{Flow of quantum communication with the surface code using multiplexed photons. 
    In the first step, a quantum state is encoded into a surface code. Each circle with a number inside denotes the physical data qubits, and the grey circles without any numbers are auxiliary qubits used for stabilizer measurements.
    For the second step in a quantum multiplexing scenario, one assigns each physical qubit of the codeword to single photons using an assignment strategy. For instance, in this figure, two components of the time-bin DOF in each photon are used so that each photon can encode two qubits. There is freedom in which qubit is assigned to which photon, so it is necessary to define a function mapping qubits to photons. We call this function the assignment strategy. Here, the colors of the qubits indicate the photon to which it is assigned based on the assignment strategy. Then, the encoded photons pass through a lossy channel. Here, we assume that we know which photons have been lost during the transmission (erasure channel). If a photon has been lost, all the qubits assigned to this photon have been lost. Finally, we demultiplex the received photons and decode it to a codeword of the surface code using the peeling decoder~\cite{delfosse2020linear} and a correction method for erasure errors as described in Sec.~\ref{sec_ER}.}
    \label{fig:sc-erasure}
\end{figure*}

We describe the steps to perform error-corrected quantum communication over a multiplexed erasure channel using the surface code as an example, as illustrated in Fig.~\ref{fig:sc-erasure}. For the first step, the sender prepares an encoded quantum state. Then, in the second step, the sender assigns and converts physical data qubits to photons. In the conventional case, different qubits are assigned to different photons, whereas when using quantum multiplexing, different qubits can be assigned to the same photon.
In the instance of Fig.~\ref{fig:sc-erasure}, qubits $0$ and $6$ are assigned to photon $0$, qubits $1$ and $7$ are assigned to photon $1$, etc. We will discuss the optimal assignment strategy later.

For the third step, a codeword passes through an optical channel which has the possibility of erasure errors. During the transmission, some photons can be lost, causing the loss of all qubits assigned to these photons as well. For instance, when photon $1$ is lost, qubits $1$ and $7$ will also be lost as shown in Fig.~\ref{fig:sc-erasure}, resulting in an erasure error. In the fourth step, the receiver reconstructs an incomplete codeword missing some qubits from the non-erased qubits assigned to the remaining photons. For the final step, the receiver replaces the missing qubits with mixed states as explained in Sec.~\ref{sec_ER}. Effectively, this results in a codeword with random Pauli errors on some qubits. Next, the decoding algorithm estimates the errors, and the receiver performs a correction. In the second half of this paper, we examine the performance of this communication procedure via numerical simulation.

\section{Multiplexed quantum communication with error-correcting codes}
\label{sec_mqcwithecc}
We propose three different scenarios in which multiplexing is used to enhance the efficiency of quantum communication. In each case, $m$ qubits are encoded into each photon, and we compare them to the case of transmitting a codeword of a given code $C$ without multiplexing ($m=1$). The three scenarios are:
\renewcommand{\labelenumi}{(\Alph{enumi}).}
\begin{enumerate}
    \item $m$ codewords of $C$ are transmitted using the same number of photons as the $m=1$ case;
    \item an $m$-times larger code from the same family as $C$ is transmitted using the same number of photons as the $m=1$ case;
    \item a codeword of $C$ is transmitted using $m$-times fewer photons than the $m=1$ case.
\end{enumerate}

Examples for the case of the surface code are shown in Table.~\ref{table:scenarios}, where the parameters of this code are given as $[\![2d^2,2,d]\!]$, with $d$ being the code distance. Let us now explore each scenario in turn.

\begin{table*}[htbp]
\scalebox{0.75}[0.75]{
\begin{tabular}{|l|c|c|c|c|c}
\cline{1-5}
 Scenarios   
  & 
    \begin{tabular}{c}
        \begin{minipage}[b]{0.1\linewidth}
        \vspace{-40pt}
        \scalebox{0.5}{\includegraphics[]{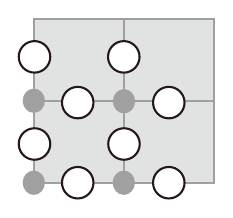}}
        \end{minipage}
    \\without multiplexing
    \end{tabular} 
 & 
 \begin{minipage}[b]{0.13\linewidth}
 \centering
 \scalebox{0.4}{\includegraphics[]{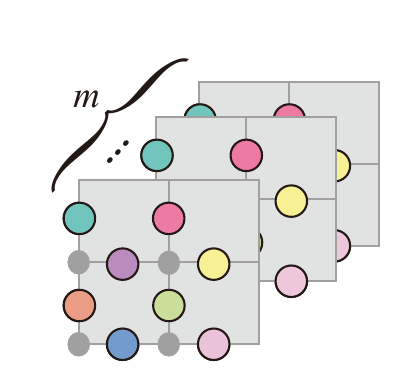}}
 \end{minipage}
 (A)
 &
  \begin{minipage}[b]{0.16\linewidth}
 \centering
 \scalebox{0.3}{\includegraphics[]{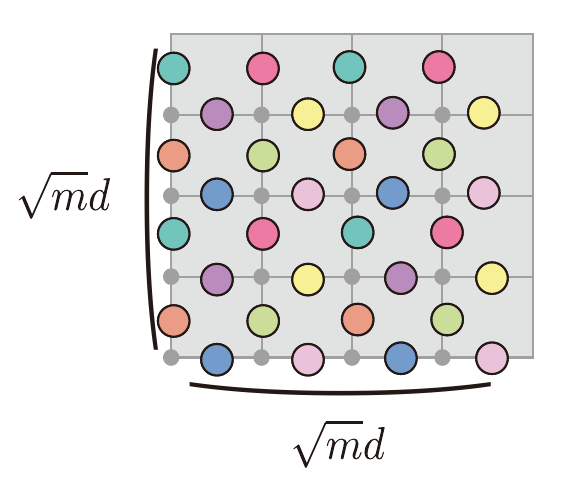}}
 \end{minipage}
 (B)
 &
  \begin{minipage}[b]{0.13\linewidth}
 \centering
 \scalebox{0.5}{\includegraphics[]{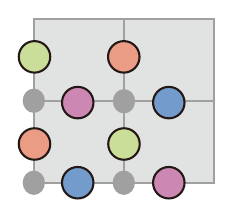}}
 \end{minipage}
 (C)
 &
 \\ \cline{1-5}
 Code parameters &$[\![2d^2,2,d]\!]$ & $[\![2d^2,2,d]\!]$ &$[\![2md^2,2,\red{\sqrt{m} d}]\!]$ &$[\![2d^2,2,d]\!]$ \\ \cline{1-5}
 Number of Codes &$1$ & \red{$m$}        & $1$    & $1$ & \\ \cline{1-5}
 Number of Data Qubits & $2d^2$ & $2md^2$  & $2md^2$  & $2d^2$ &  \\ \cline{1-5}
 Number of Photons& $2d^2$ & $2d^2$  & $2d^2$ & \red{$\lfloor 2d^2/m \rfloor$} &  \\ \cline{1-5}
 Logical Error Rate & 
 -
 &
 \begin{tabular}{c}
 Same as without\\
 quantum multiplexing
 \end{tabular}
&
Affected by correlation
 &  
Affected by correlation
 \\ \cline{1-5}
\end{tabular}}
\caption{Comparison of surface code communication without multiplexing and three scenarios with multiplexing. Red fonts show parameters that are improved by multiplexing. The case without multiplexing requires one qubit per photon. (A) The first scenario is only applicable when sending multiple codewords. This enables one to send more codewords with the same number of photons, drastically improving the channel's throughput. (B) The second scenario sends the same number of codewords but uses a larger code, thus improving the error tolerance. (C) The third scenario sends the same codeword with fewer photons, which also significantly improves the channel's throughput. The number of photons required in scenario (C) is $\lfloor 2d^2/m \rfloor$, where $\lfloor x \rfloor$ is the floor function of $x$.}
\label{table:scenarios}
\end{table*}

\subsection*{(A) Sending \textit{$m$} different codewords}
In the first scenario, the multiplexed photons are used to encode $m$ codewords from $m$ independent copies of the same code. The logical throughput of the channel increases $m$ fold over the no-multiplexing case. One can assign qubits to photons so that each photon contains one qubit from a codeword of the distinct codes. The qubits from different codewords are correlated, but there is no correlation among the qubits in a fixed code. This correlation does not affect the logical error rate of the individual codes. 

\subsection*{(B) Sending \texorpdfstring{$m$}\,-times bigger codewords}
In the second scenario, a larger number of qubits is used to encode a single codeword from a code in the same family with an $m$-fold longer length. If this scenario is applied to the surface code, it achieves $\sqrt{m}$ times larger distance than the no-multiplexing case ($m=1$).

\begin{figure}[ht]
    \centering
    \includegraphics[width = 1 \columnwidth]{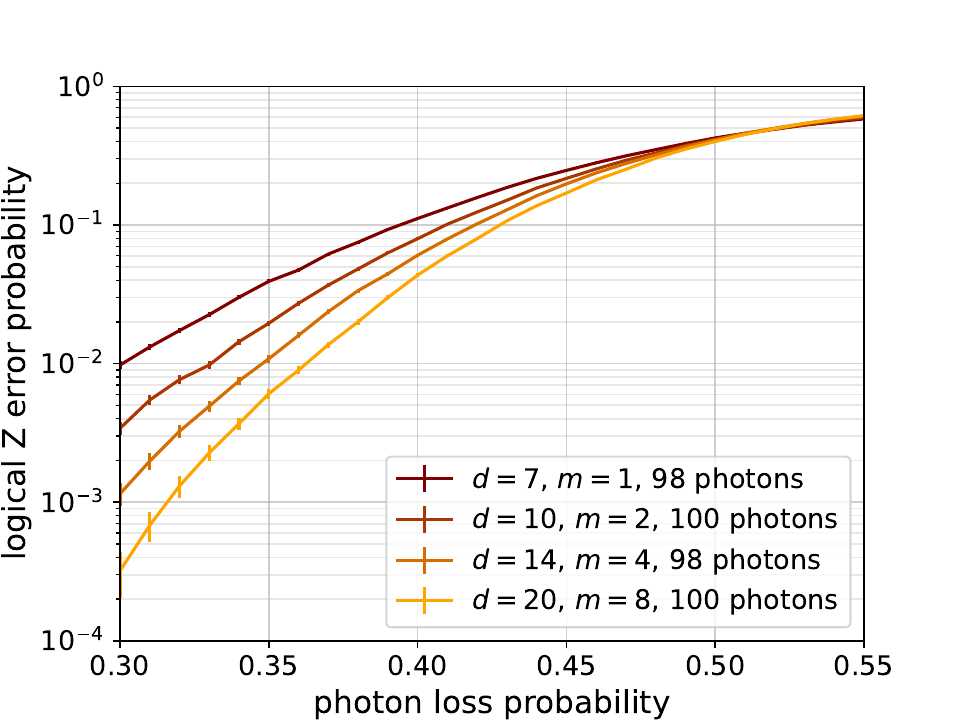}
\caption{Performance of $[\![2d^2,2,d]\!]$ surface codes in scenario (B) with about $100$ photons. Each curve shows the case with different code sizes and the number of qubits encoded in each photon. The logical error rate can be reduced by increasing the number of qubits per photon $m$ and the code distance $d$.}
    \label{fig:7-10}
\end{figure}

Fig.~\ref{fig:7-10} shows a Monte Carlo simulation of the logical $Z$ error rate for this scenario for the surface code. We use surface codes only on a 2$D$-torus (i.e. toric codes) in this paper.
To determine whether a logical $Z$ error occurred, we checked whether the errors left after the decoding process were anti-commutative with a logical $X$ operator of any logical qubit in the codeword.
As logical $Y$ errors can be viewed as combinations of logical $X$ and $Z$ errors, we also count these as logical $Z$ errors for the purpose of these plots.
Each data point in the simulation is obtained from $10^5$ shots, and the error bar is given by the Agresti–Coull interval~\cite{agresti1998approximate}. For simplicity, we assign qubits to multiplexed photons uniformly at random in this section. This scenario introduces correlations in errors between the qubits in the code, which may degrade the performance. However, if $m$ is sufficiently small relative to the code size, the benefit gained by increasing the code size is more significant. All the programs we used to simulate multiplexed quantum communication with surface codes are available here~\cite{Nishio_C_implementation_of_2024}.
The logical error rate significantly decreases as the code size and $m$ increase. 

Readers may wonder why the logical $Z$ error rate converges to $0.75$, even though for a loss probability of $1$ the code will be completely erased. This is due to the fact that, following such a complete erasure, the recovery procedure will reinitialize the code in a random state subject to one of the logical operators $II,ZI,IZ,ZZ$ with equal probability, and so $25\%$ of the time we do not observe a logical $Z$ error. Importantly, this finite recovery probability is observed because, in this case, we care only about the presence/absence of logical $Z$ operators, and it does not imply that an arbitrary quantum state can be recovered.

\subsection*{(C) Sending original codewords with fewer photons}
\begin{figure}[htbp]
    \centering
    \includegraphics[width = 1 \columnwidth]{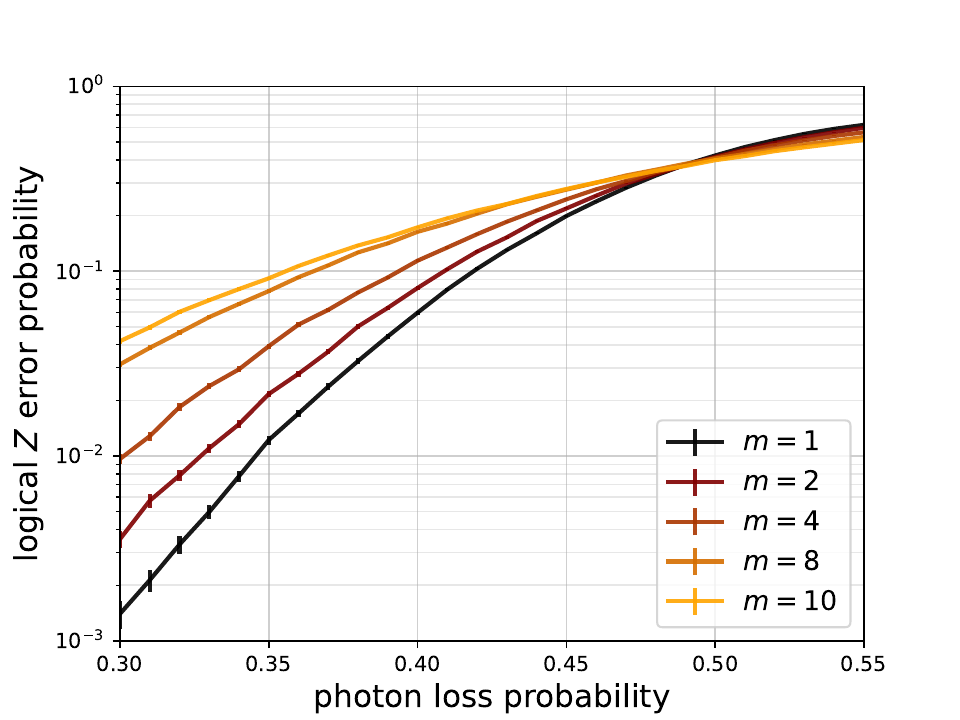}
    \caption{Scenario (C) multiplexing performance for the $[\![200,2,10]\!]$ surface code with multiplexing using different values of $m$ (the number of qubits per photon). The assignment of qubits to photons is uniformly random. Increasing $m$ allows code words to be transmitted with fewer photons, but the logical error rate increases because multiple qubits in the same photon have strongly correlated errors.}
    \label{fig:m}
\end{figure}

In the third scenario, a smaller number of photons are used to encode a single codeword. The code parameters are the same as the case without multiplexing.
It has no restriction on the number of codewords and can improve the efficiency of surface code communication in general. This method introduces a correlation to the errors. Fig.~\ref{fig:m} shows this scenario's logical $Z$ error rate versus the photon loss probability for different values of $m$. It shows that as $m$ increases, the logical error rate also increases.

Although the number of photons is reduced compared to the no-multiplexing case, the effects of the correlated errors can be very detrimental to the performance of such a system.
Ideally, these detrimental effects can be reduced by strategically assigning qubits to the multiplexed photons.
In the next subsection, we explore five different strategies for qubit assignment.

\subsection{Limitation and fault-tolerance of multiplexed communication}
Generating time-bin multiplexed photons requires linear optical elements such as polarizing beam splitters, delay lines, wave plates, and optical switches, as shown in Fig.~\ref{fig:qm_encoder}~(A).  In particular, the latter suffer from loss errors that are detrimental to the overall performance of the system. In \cite{piparo2019quantum}, the authors analyze the impact of imperfect optical switches on the performance of a purification protocol. They show that, for an optical switch efficiency equal to 99\%, adding one time-bin mode to a polarized photon only slightly reduces the performance of the system. However, when several optical switches are in use to encode multiple time-bin modes, much higher efficiencies are required. Although not fully achieved with the current devices, future high-efficiency optical switches are required for several quantum technologies.

While it is possible for multiplexing to lead to correlated errors between qubits stored in the same photon, we do not expect this to compromise fault tolerance except in the case where these qubits correspond to the support of a logical operator of the code, and the majority of our assignment strategies are designed to prevent this. More generally, if we require that the number of qubits per photon is a) independent of the size of the code and b) smaller than the code distance, then these correlated errors can, at most, result in a constant-factor reduction of the distance, and so long as the non-multiplexed procedure is fault-tolerant, the multiplexed procedure will be fault-tolerant as well.

\section{Quantum Communication with Multiplexed Surface Codes}
\label{sec_SC}
\subsection{Assignment Strategies for Surface Codes}
\label{sec_AS}
In this section, we describe five strategies for assigning qubits to photons which take advantage of multiplexing, and then evaluate the impact of these strategies on communication performance. These strategies assume that each photon contains a fixed number of qubits $m$ and can be applied in both scenarios \textbf{(B)} and \textbf{(C)}. Because \textbf{(B)} can be regarded as a scaled version of scenario \textbf{(C)}, we will focus specifically on surface code communication using the latter case, where we send the original code word with $\lfloor 2d^2/m\rfloor$ photons.\\

\begin{figure}[ht]
  \begin{minipage}[b]{0.49\linewidth}
    \centering
    \includegraphics[keepaspectratio, scale=0.7]{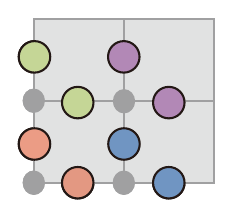}
    \\(A)
  \end{minipage}
  \begin{minipage}[b]{0.49\linewidth}
    \centering
    \includegraphics[keepaspectratio, scale=0.7]{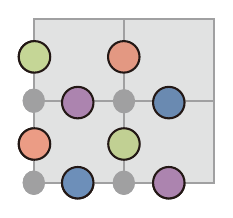}
    \\(B)
  \end{minipage}
  \caption{Examples showing possible assignments of qubits to photons. Each colored circle denotes a qubit, and the color indicates the photon to which the qubit is assigned. Each non-colored (gray) small circle denotes auxiliary qubits for stabilizer measurement. Strategy i (shown in (A)) minimizes the distance between qubits in the same photon, while strategy ii (shown in (B)) maximizes this distance. Note that this code is defined on the torus represented as a lattice with periodic boundary conditions.}
  \label{fig:minmax}
\end{figure}

\textbf{Strategy i and ii: pair with minimum and maximum distance}
Strategies i and ii are applicable to the case with $m = 2$. Strategy i assigns the nearest neighbor pair of qubits to the same photon, which forms an L-shape in the 2D lattice of the surface code as shown in the example of Fig.~\ref{fig:minmax}(A). This minimizes the Manhattan distance in the lattice between the qubits in the same photon.
Strategy ii assigns the qubit at coordinates $(i,j)$ on the lattice and the qubit at coordinates $(i+d/2 - 1 \,\textrm{mod} \,d, j + d/2 - 1 \,\textrm{mod} \,d)$ to the same photon. An example is shown in Fig.~\ref{fig:minmax}(B). This is the arrangement that maximizes the Manhattan distance, in contrast to strategy i.

\textbf{Strategy iii: random} Strategy iii is a method in which qubits are selected uniformly at random and assigned to photons. 

\textbf{Strategy iv: random + threshold} 
Strategy iv is designed to increase the separation between qubits assigned to the same photon while exploiting randomness. The strategy works by randomly selecting qubits and accepting them as the set for a photon only if the distance is greater than a certain threshold. If no suitable set of qubits can be found, then the threshold value is reduced.

\begin{figure}[htbp]
\begin{tabular}{ccc}
     \begin{minipage}[b]{0.3\linewidth}
     \centering
     \scalebox{0.35}{\includegraphics[]{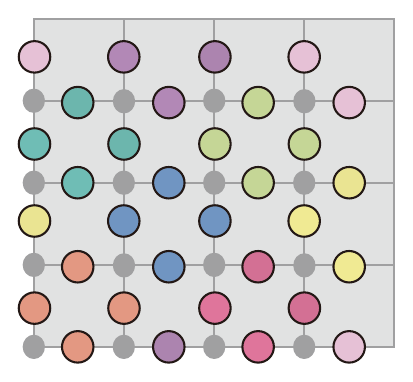}}
     \end{minipage}
    &
     \begin{minipage}[b]{0.3\linewidth}
     \centering
     \scalebox{0.35}{\includegraphics[]{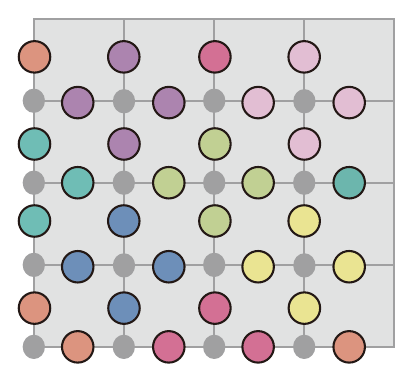}}
     \end{minipage}
    &
     \begin{minipage}[b]{0.3\linewidth}
     \centering
     \scalebox{0.35}{\includegraphics[]{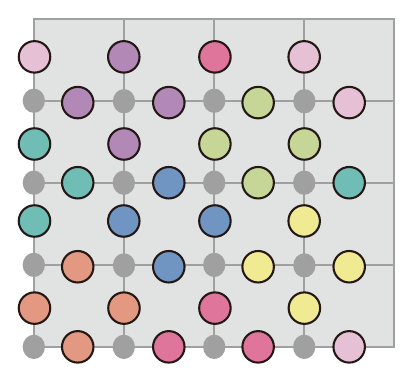}}
     \end{minipage}
    \\
 {\bf $Z$-stabilizer} & {\bf $X$-stabilizer} & {\bf Mixed}\\ 
\end{tabular}
\caption{Examples of the stabilizer-based photon assignment strategy for a surface code on a $4\times4$ lattice. Edges representing qubits in the lattice are marked with colored nodes indicating photon assignment. In this lattice picture, the qubit support of $Z$-type stabilizer generators corresponds to squares, and of $X$-type stabilizer generators corresponds to crosses. Each photon in the stabilizer assignment strategy represents the qubit-support of one of these stabilizers. 
}
\label{fig:surface_stabilizer}
\end{figure}

\textbf{Strategy v: stabilizer}
Strategy v assigns the qubit support of stabilizer generators to the same photon. Realizations of this assignment strategy on a $4\times4$ surface code are shown in Fig.~\ref{fig:surface_stabilizer}.

We describe the details and the motivation of each strategy in Appendix~\ref{app:strategies_for_sc}.

\begin{figure*}
\begin{tabular}{cc}
     \begin{minipage}[b]{0.5\linewidth}
     \centering
     \includegraphics[width = 1 \columnwidth]{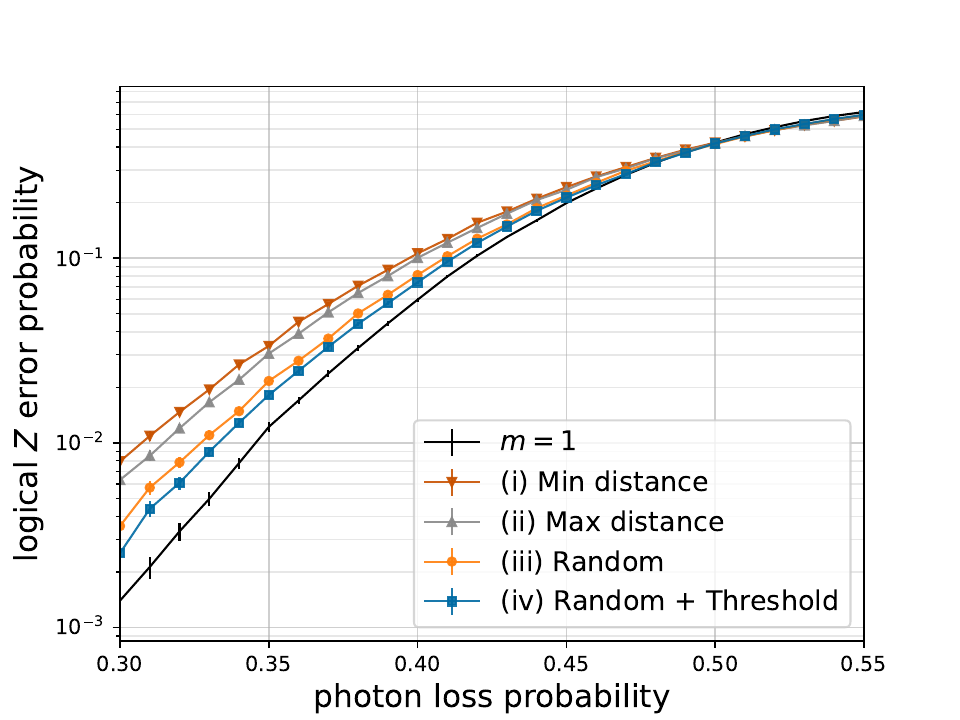}
     \end{minipage}
    &
     \begin{minipage}[b]{0.5\linewidth}
     \centering
     \includegraphics[width = 1 \columnwidth]{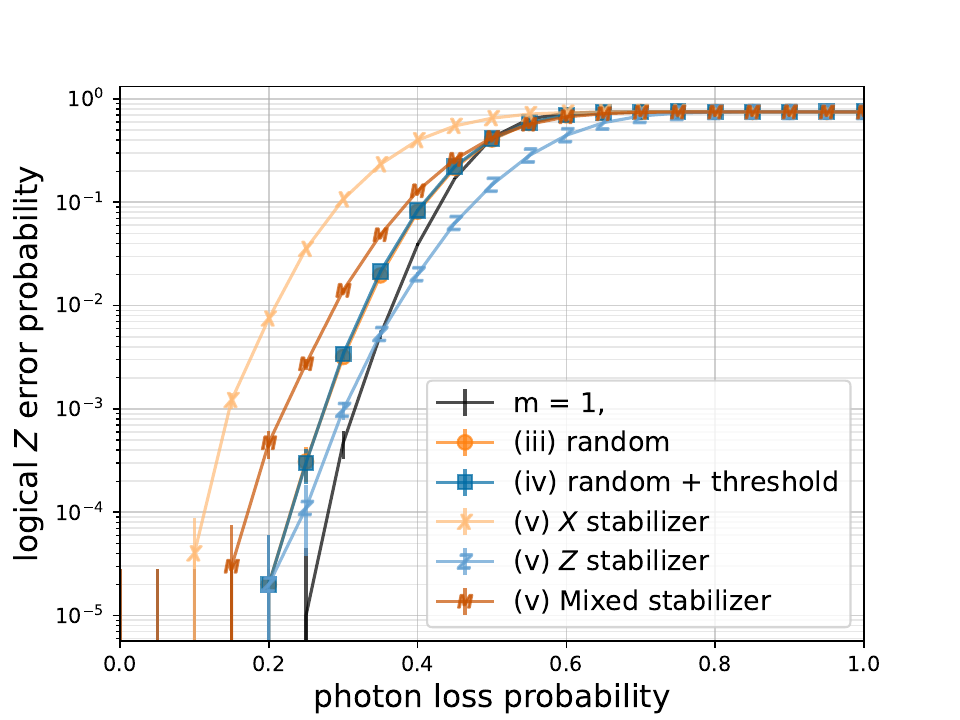}
     \end{minipage}
    \\
 (A) & (B)\\ 
\end{tabular}
\caption{Comparison of multiplexing photon-assignment strategies for surface codes. Logical $Z$ error rate versus photon loss probability. The black curve shows the case without multiplexing. (A) The code parameters are $[\![200,2,10]\!]$ and $m=2$. The gray/brown curve shows the case for the assignment strategy for minimizing (strategy i) / maximizing (strategy ii) the distance between a pair of qubits in the same photon. The orange curve shows the case for uniform randomness (strategy iii), and the blue line shows random + threshold (strategy iv), based on Algorithm~\ref{alg:randomplusthreshold}. (B) The code parameters are $[\![288,2,12]\!]$ and $m=4$. The $Z$ stabilizer-based assignment (light blue curve) outperformed the $X$ stabilizer-based assignment (light orange curve) for logical $Z$ errors. The mixed stabilizer-based assignment strategy performs between $X$ and $Z$. Strategy iii (orange circle) and iv (blue square) outperform the other assignment strategies when the photon loss probability is low.}
\label{fig:strategies_for_sc}
\end{figure*}
\subsection{Performance of the assignment strategies}
Here, we show the performances of these strategies observed in numerical simulations.

Fig.~\ref{fig:strategies_for_sc}~(A) shows the performance of strategies i through iv, which are applicable to the case of $m=2$. The distance-maximizing strategy (grey) outperforms the distance-minimizing strategy (brown). Logical errors in the surface code correspond to errors covering a longitude or meridian curve on the torus (a vertical or horizontal closed loop in the periodic lattice). When decoding erasure errors, logical errors can only occur when the qubit-support of one of these vertical or horizontal loops is entirely erased. When adjacent qubits in the lattice are erased, as in the case with the distance-minimizing photon assignment strategy, clusters of errors are more likely to cover such loops in the torus. Hence, it is not surprising that the distance-maximizing strategy outperforms the distance-minimizing strategy in our numerical simulations. We also see that the strategies with randomness (iii and iv) outperform the deterministic ones (i and ii). In particular, strategy iv outperformed the other strategies, although there was an increase in logical $Z$-error probability compared to no multiplexing.

Next, we compare the logical $Z$ error rates of strategies iii, iv, and v with $m=4$ in Fig.~\ref{fig:strategies_for_sc}~(B). Assignment strategies based on one type of stabilizer create a bias in observed logical error rates. Strategy v can be generalized to any stabilizer code, and the assignment strategy based only on the support of $X$ or $Z$ stabilizers will increase the error rate of one of $X$ or $Z$ and decrease the other. This result implies that stabilizer-based assignment may be useful in quantum error correction codes with different $X$ and $Z$ distances. 

Both Fig.~\ref{fig:strategies_for_sc}~(A) and (B) showed that the strategy iv random + threshold outperformed the other strategies. Maximizing the distance between qubits while also introducing randomness gives the largest boost in performance against the occurrence of logical errors. Note that no assignment strategy does better than the case with $m=1$ where no multiplexing is used.

\begin{figure}[ht]
    \centering
    \includegraphics[width = 1 \columnwidth]{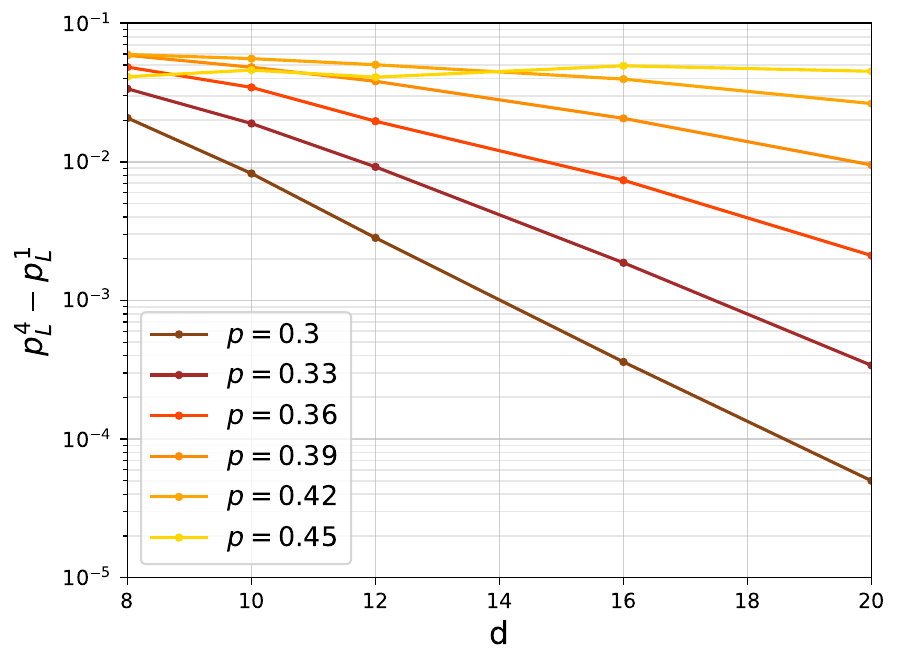}
    \caption{Difference of logical $Z$ error rates for $m=4$ ($p_{L}^4$) and $m=1$ ($p_{L}^1$) for various photon loss probabilities ($p$). For low $p$ ($0.3$ \textasciitilde \,$0.42$), the gap decreases to $0$ as $d$ increases.}
    \label{fig:8_16}
\end{figure}

We also analyzed the difference in the performance between cases with multiplexing ($m=4$) and without it ($m=1$), as shown in Fig.~\ref{fig:8_16}. When physical error rates are low, this difference decreases with increasing code distance, suggesting that the downsides of multiplexing are less significant in larger codes. 

\begin{figure*}[htbp]
    \centering
    \begin{tabular}{cc}
         \begin{minipage}[b]{0.5\linewidth}
         \centering
         \includegraphics[width = 1 \columnwidth]{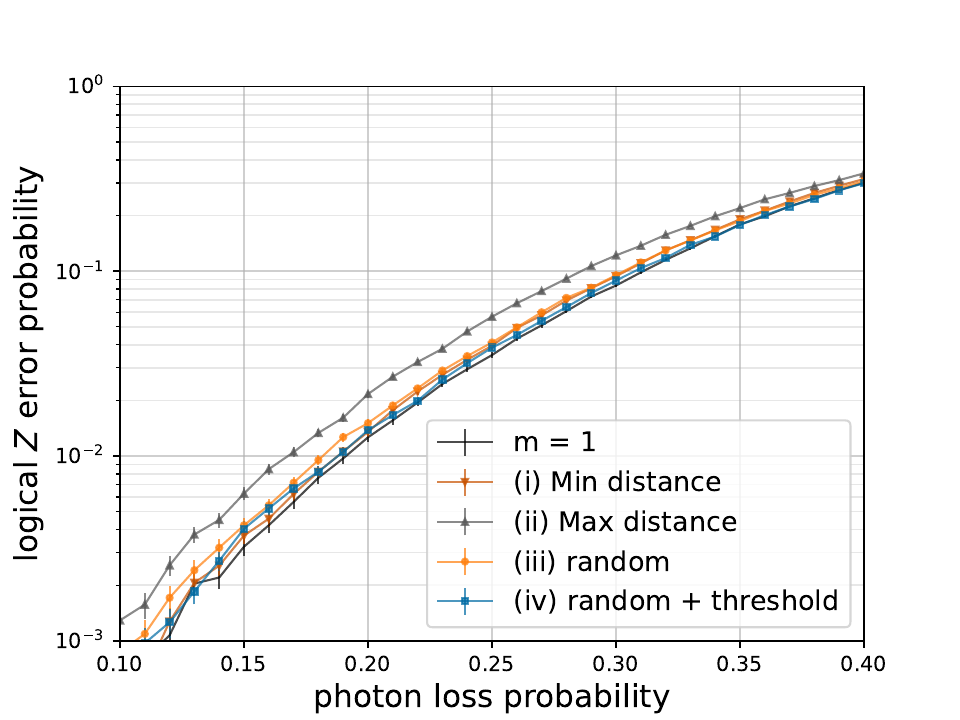}
         \end{minipage}
        &
         \begin{minipage}[b]{0.5\linewidth}
         \centering
         \includegraphics[width = 1 \columnwidth]{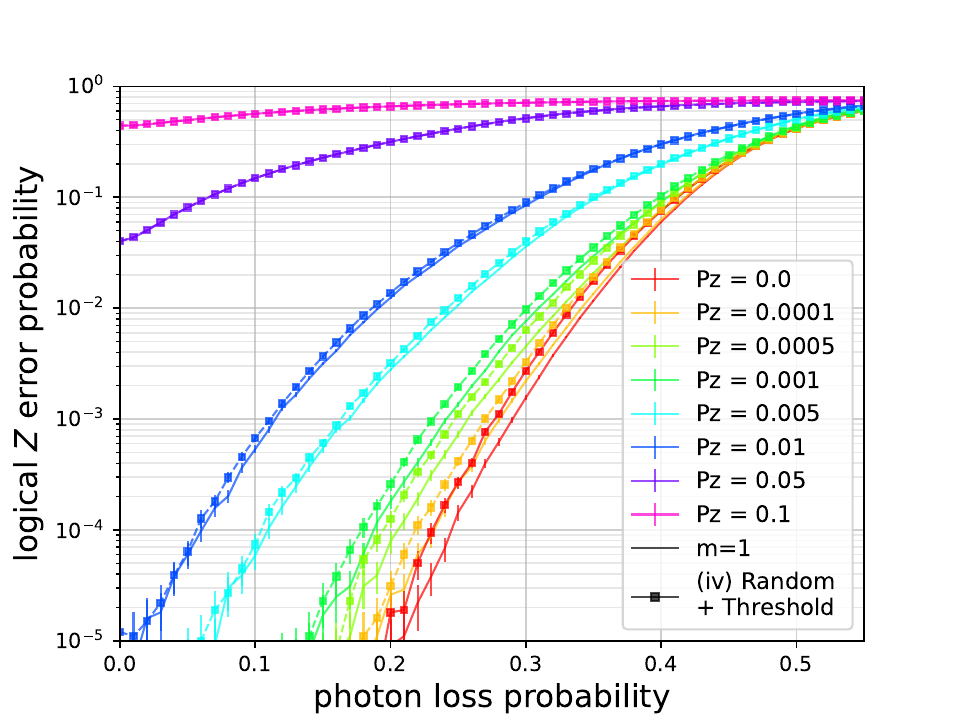}
         \end{minipage}
        \\
     (A) & (B)\\ 
    \end{tabular}
    \caption{(A) Comparison of $m=2$ photon-assignment strategies on the $[\![200,2,10]\!]$ surface code with the combined error model. The combined error consists of erasure errors and $1$\% Pauli $Z$ errors occurring before and after the multiplexed communication. (B) The logical error rate for $m=1$ and $m=2$ (strategy (iv)) with different probability of Pauli $Z$ errors $P_Z$. $10^6$ samples are taken. Color indicates $P_Z$, and solid and dashed lines indicate $m=1$ and $m=2$, respectively.}
    \label{fig:strategies_for_sc_combined}
\end{figure*}

To assess the impact of multiplexing on the logical error rate in more realistic use cases, we evaluated the performance in a mixed error model where both Pauli errors and photon losses are present. We performed a simulation with the following flow.\\
(1) Random Pauli errors on sender's processor $\xrightarrow{\rm transfer\,out\,and\,send}$ (2) Erasure errors on (multiplexed) photonic states $\xrightarrow{\rm transfer\,in}$ (3) Random Pauli errors on receiver's processor $\rightarrow$ (4) Quantum nondemolition measurement to detect erasure errors $\rightarrow$ (5) Replace erased qubits with mixed state $\rightarrow$ (6) Stabilizer measurement which maps erasure errors to random Pauli errors $\rightarrow$ (7) Decoding procedure. Here, ``transfer out'' is a procedure to encode qubits into a photon, and ``transfer in'' is the opposite procedure.\\

Note that random Pauli errors in (1) have a smaller effect on the logical error rate than in (3) because some erasure errors in (2) will eliminate the effect of (1) on the erased qubits.
In this setting, there are Pauli errors only before and after the communication part, and there are no Pauli errors on the multiplexed channel. The behavior of Pauli errors on the multiplexed state is not as obvious as the single-qubit Pauli errors because, for such a case, the Pauli errors may be on qudits, and it can have a larger effect than single-qubit Pauli errors. This qudit Pauli on the multiplexed channel is complex, so we will leave this problem to future work.\\

Note that we have implemented and used the Union-find decoder~\cite{delfosse2021almost} for this new simulation, which can handle the combined errors. Fig.~\ref{fig:strategies_for_sc_combined}~(A) shows the logical error rate with the combined error for different strategies. The performance of strategies (i) and (ii) is reversed compared to the case without Pauli error in Fig.~\ref{fig:strategies_for_sc}~(A). The strategy (iv) achieves a logical error rate relatively closer to $m = 1$ than (iii) does. 
Fig.~\ref{fig:strategies_for_sc_combined}~(B) shows the logical error rate versus photon loss probability with different Pauli error rates $P_Z$ for $m=1$ and $m=2$ (strategy (iv)). The $m=1$ case outperforms strategy (iv) for low values of $P_Z$, but for high values, strategy (iv) is close to no multiplexing and outperforms only slightly for some loss error rates. These indicate that the method of choosing an appropriate assignment strategy in more practical cases may require further consideration.\par
Note that the threshold~\cite{stace2009thresholds} for the surface code with Union-find decoder is $9.9$~\%~\cite{delfosse2021almost}, but here we have random Pauli $Z$ errors occurring before and after the communication process, and so we expect the threshold to be roughly half the original threshold.\\\par
\clearpage
\section{Quantum Communication with Multiplexed Hypergraph Product Codes}
\label{sec_HGP}
\subsection{Hypergraph Product Code Structure}
In addition to our exploration of the surface code, we also consider the use of multiplexing with hypergraph product (HGP) codes~\cite{tillich2013quantum}, of which surface codes are a special case. HGP codes are a special class of CSS code defined using any two classical linear codes. They are of particular interest because they can have an asymptotically finite rate as the code length increases (in contrast with the surface code, which has a rate approaching $0$) and distance proportional to the minimum distance of the classical codes; in the best case, this is proportional to the square root of the quantum code length. They are also considered practical candidates for FTQC codes. 

Given classical parity check matrices $H_1$ and $H_2$ with sizes $r_1\times n_1$ and $r_2\times n_2$, respectively, we may define the matrices $H_X$ and $H_Z$ of a CSS code via the formulas
\begin{eqnarray}
    H_X&=&(H_1\otimes I_{n_2}|I_{r_1}\otimes H_2^T)\label{eq:HGP_Hx}\\
    H_Z&=&(I_{n_1}\otimes H_2|H_1^T\otimes I_{r_2})\label{eq:HGP_Hz},
\end{eqnarray}
where $I_{n_1}$, $I_{n_2}$, $I_{r_1}$, $I_{r_2}$ are identity matrices with sizes $n_1$, $n_2$, $r_1$, and $r_2$, respectively. These matrices satisfy the condition $H_XH_Z^T=0$ by construction and hence define a valid CSS code $\text{HGP}(H_1,H_2)$. When $H_1$ and $H_2$ are low-density parity checks (LDPC), $H_X$ and $H_Z$ will also be LDPC. The sizes of $H_X$ and $H_Z$ are determined by the sizes of the input classical matrices according to the formulas
\begin{eqnarray}
    H_X&=&[r_1n_2\times(n_1n_2+r_1r_2)]\label{eq:HGP_Hx_size}\\
    H_Z&=&[r_2n_1\times(n_1n_2+r_1r_2)]\label{eq:HGP_Hz_size}.
\end{eqnarray}
These both simplify to $rn\times(n^2+r^2)$ in the special case where $r_1=r_2=r$ and $n_1=n_2=n$.

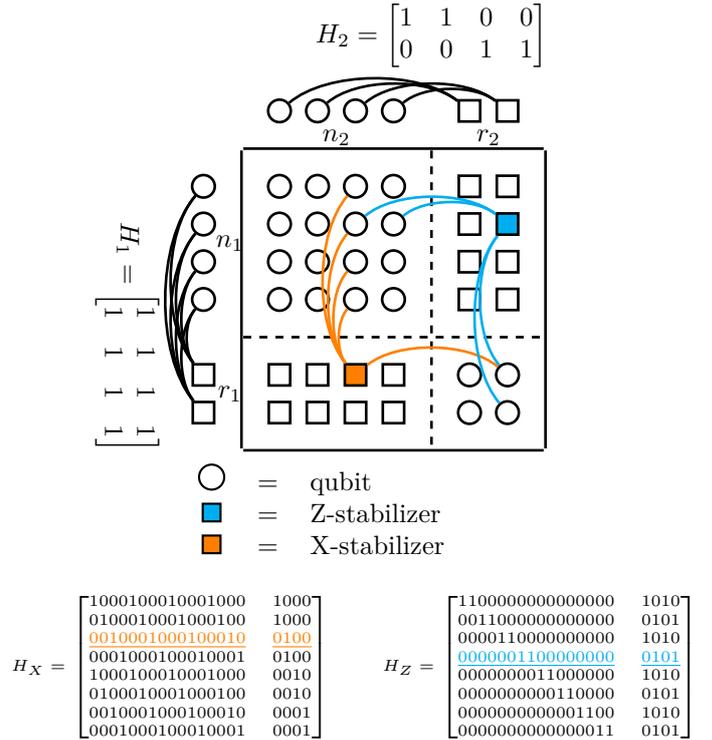
\begin{figure}[t]
\begin{center}
{\small
\begin{tikzpicture}[baseline]
\node (nw) [inner sep=0] at (-5/2,5/2) {};
\node (n) [inner sep=0] at (0,5/2) {};
\node (ne) [inner sep=0] at (3/2,5/2) {};
\node (w) [inner sep=0] at (-5/2,0) {};
\node (c) [inner sep=0] at (0,0) {};
\node (e) [inner sep=0] at (3/2,0) {};
\node (sw) [inner sep=0] at (-5/2,-3/2) {};
\node (s) [inner sep=0] at (0,-3/2) {};
\node (se) [inner sep=0] at (3/2,-3/2) {};

\path (nw) edge[line width=1,opacity=1] (ne);
\path (w) edge[line width=1,dashed,opacity=1] (e);
\path (sw) edge[line width=1,opacity=1] (se);
\path (nw) edge[line width=1,opacity=1] (sw);
\path (n) edge[line width=1,dashed,opacity=1] (s);
\path (ne) edge[line width=1,opacity=1] (se);

\node(n1) at (-5.3/2,2.5/2) {\rotatebox{0}{$n_1$}};
\node(n2) at (-2.5/2,5.3/2) {\rotatebox{0}{$n_2$}};
\node(r1) at (-5.3/2,-1.5/2) {\rotatebox{0}{$r_1$}};
\node(r2) at (1.5/2,5.3/2) {\rotatebox{0}{$r_2$}};

\node (-44) [circle,draw,inner sep=3pt,line width=1,opacity=1] at (-4/2,4/2) {};
\node (-34) [circle,draw,inner sep=3pt,line width=1,opacity=1] at (-3/2,4/2) {};
\node (-24) [circle,draw,inner sep=3pt,line width=1,opacity=1] at (-2/2,4/2) {};
\node (-14) [circle,draw,inner sep=3pt,line width=1,opacity=1] at (-1/2,4/2) {};
\node (-43) [circle,draw,inner sep=3pt,line width=1,opacity=1] at (-4/2,3/2) {};
\node (-33) [circle,draw,inner sep=3pt,line width=1,opacity=1] at (-3/2,3/2) {};
\node (-23) [circle,draw,inner sep=3pt,line width=1,opacity=1] at (-2/2,3/2) {};
\node (-13) [circle,draw,inner sep=3pt,line width=1,opacity=1] at (-1/2,3/2) {};
\node (-42) [circle,draw,inner sep=3pt,line width=1,opacity=1] at (-4/2,2/2) {};
\node (-32) [circle,draw,inner sep=3pt,line width=1,opacity=1] at (-3/2,2/2) {};
\node (-22) [circle,draw,inner sep=3pt,line width=1,opacity=1] at (-2/2,2/2) {};
\node (-12) [circle,draw,inner sep=3pt,line width=1,opacity=1] at (-1/2,2/2) {};
\node (-41) [circle,draw,inner sep=3pt,line width=1,opacity=1] at (-4/2,1/2) {};
\node (-31) [circle,draw,inner sep=3pt,line width=1,opacity=1] at (-3/2,1/2) {};
\node (-21) [circle,draw,inner sep=3pt,line width=1,opacity=1] at (-2/2,1/2) {};
\node (-11) [circle,draw,inner sep=3pt,line width=1,opacity=1] at (-1/2,1/2) {};

\node (14) [rectangle,draw,inner sep=4pt,line width=1,opacity=1] at (1/2,4/2) {};
\node (24) [rectangle,draw,inner sep=4pt,line width=1,opacity=1] at (2/2,4/2) {};
\node (13) [rectangle,draw,inner sep=4pt,line width=1,opacity=1] at (1/2,3/2) {};
\node (23) [rectangle,draw,inner sep=4pt,line width=1,opacity=1,fill=cyan] at (2/2,3/2) {};
\node (12) [rectangle,draw,inner sep=4pt,line width=1,opacity=1] at (1/2,2/2) {};
\node (22) [rectangle,draw,inner sep=4pt,line width=1,opacity=1] at (2/2,2/2) {};
\node (11) [rectangle,draw,inner sep=4pt,line width=1,opacity=1] at (1/2,1/2) {};
\node (21) [rectangle,draw,inner sep=4pt,line width=1,opacity=1] at (2/2,1/2) {};

\node (-4-1) [rectangle,draw,inner sep=4pt,line width=1,opacity=1] at (-4/2,-1/2) {};
\node (-3-1) [rectangle,draw,inner sep=4pt,line width=1,opacity=1] at (-3/2,-1/2) {};
\node (-2-1) [rectangle,draw,inner sep=4pt,line width=1,opacity=1,fill=orange] at (-2/2,-1/2) {};
\node (-1-1) [rectangle,draw,inner sep=4pt,line width=1,opacity=1] at (-1/2,-1/2) {};
\node (-4-2) [rectangle,draw,inner sep=4pt,line width=1,opacity=1] at (-4/2,-2/2) {};
\node (-3-2) [rectangle,draw,inner sep=4pt,line width=1,opacity=1] at (-3/2,-2/2) {};
\node (-2-2) [rectangle,draw,inner sep=4pt,line width=1,opacity=1] at (-2/2,-2/2) {};
\node (-1-2) [rectangle,draw,inner sep=4pt,line width=1,opacity=1] at (-1/2,-2/2) {};

\node (1-1) [circle,draw,inner sep=3pt,line width=1,opacity=1] at (1/2,-1/2) {};
\node (2-1) [circle,draw,inner sep=3pt,line width=1,opacity=1] at (2/2,-1/2) {};
\node (1-2) [circle,draw,inner sep=3pt,line width=1,opacity=1] at (1/2,-2/2) {};
\node (2-2) [circle,draw,inner sep=3pt,line width=1,opacity=1] at (2/2,-2/2) {};

\node(H2) [] at (0,8/2) {$H_2=\begin{bmatrix}1&1&0&0\\0&0&1&1\end{bmatrix}$};
\node (-46) [circle,draw,inner sep=3pt,line width=1,opacity=1] at (-4/2,6/2) {};
\node (-36) [circle,draw,inner sep=3pt,line width=1,opacity=1] at (-3/2,6/2) {};
\node (-26) [circle,draw,inner sep=3pt,line width=1,opacity=1] at (-2/2,6/2) {};
\node (-16) [circle,draw,inner sep=3pt,line width=1,opacity=1] at (-1/2,6/2) {};
\node (16) [rectangle,draw,inner sep=4pt,line width=1,opacity=1] at (1/2,6/2) {};
\node (26) [rectangle,draw,inner sep=4pt,line width=1,opacity=1] at (2/2,6/2) {};
\path (-46) edge[line width=1,bend left=40,looseness=0.75] (16);
\path (-36) edge[line width=1,bend left=40,looseness=0.75] (16);
\path (-26) edge[line width=1,bend left=40,looseness=0.75] (26);
\path (-16) edge[line width=1,bend left=40,looseness=0.75] (26);

\node(H1) [] at (-8/2,0) {\rotatebox{-90}{$H_1=\begin{bmatrix}1&1&1&1\\1&1&1&1\end{bmatrix}$}};
\node (-64) [circle,draw,inner sep=3pt,line width=1,opacity=1] at (-6/2,4/2) {};
\node (-63) [circle,draw,inner sep=3pt,line width=1,opacity=1] at (-6/2,3/2) {};
\node (-62) [circle,draw,inner sep=3pt,line width=1,opacity=1] at (-6/2,2/2) {};
\node (-61) [circle,draw,inner sep=3pt,line width=1,opacity=1] at (-6/2,1/2) {};
\node (-6-1) [rectangle,draw,inner sep=4pt,line width=1,opacity=1] at (-6/2,-1/2) {};
\node (-6-2) [rectangle,draw,inner sep=4pt,line width=1,opacity=1] at (-6/2,-2/2) {};
\path (-64) edge[line width=1,bend left=-40,looseness=0.75] (-6-1);
\path (-64) edge[line width=1,bend left=-40,looseness=0.75] (-6-2);
\path (-63) edge[line width=1,bend left=-40,looseness=0.75] (-6-1);
\path (-63) edge[line width=1,bend left=-40,looseness=0.75] (-6-2);
\path (-62) edge[line width=1,bend left=-40,looseness=0.75] (-6-1);
\path (-62) edge[line width=1,bend left=-40,looseness=0.75] (-6-2);
\path (-61) edge[line width=1,bend left=-40,looseness=0.75] (-6-1);
\path (-61) edge[line width=1,bend left=-40,looseness=0.75] (-6-2);

\path (-24) edge[line width=1,bend left=-40,looseness=0.75,color=orange] (-2-1);
\path (-23) edge[line width=1,bend left=-40,looseness=0.75,color=orange] (-2-1);
\path (-22) edge[line width=1,bend left=-40,looseness=0.75,color=orange] (-2-1);
\path (-21) edge[line width=1,bend left=-40,looseness=0.75,color=orange] (-2-1);
\path (-2-1) edge[line width=1,bend left=40,looseness=0.75,color=orange] (2-1);

\path (-23) edge[line width=1,bend left=40,looseness=0.75,color=cyan] (23);
\path (-13) edge[line width=1,bend left=40,looseness=0.75,color=cyan] (23);
\path (23) edge[line width=1,bend left=-40,looseness=0.75,color=cyan] (2-1);
\path (23) edge[line width=1,bend left=-40,looseness=0.75,color=cyan] (2-2);

\end{tikzpicture}
\begin{tabular}{ccl}$\tikz{\node [circle,draw,thick] at (0,0) {};}$&=&qubit\\$\tikz{\node [rectangle,draw,thick,fill=cyan] at (0,0) {};}$&=&Z-stabilizer\\$\tikz{\node [rectangle,draw,thick,fill=orange] at (0,0) {};}$&=&X-stabilizer\end{tabular}
{\tiny
\begin{eqnarray*}
H_X=\begin{bmatrix}1000100010001000&1000\\
0100010001000100&1000\\
\color{orange}{\underline{0010001000100010}}&\color{orange}{\underline{0100}}\\
0001000100010001&0100\\
1000100010001000&0010\\
0100010001000100&0010\\
0010001000100010&0001\\
0001000100010001&0001\end{bmatrix}&&
H_Z=\begin{bmatrix}1100000000000000&1010\\
0011000000000000&0101\\
0000110000000000&1010\\
\color{cyan}{\underline{0000001100000000}}&\color{cyan}{\underline{0101}}\\
0000000011000000&1010\\
0000000000110000&0101\\
0000000000001100&1010\\
0000000000000011&0101\end{bmatrix}
\end{eqnarray*}
}
}
\end{center}
    \caption{Example of the Tanner graph for a simple HGP code $\text{HGP}(H_1,H_2)$ constructed from two classical codes with parity check matrices $H_1$ and $H_2$. This is the cartesian product of two classical Tanner graphs, and the subgraph corresponding to each row and column in the product is a copy of one of these classical Tanner graphs. This product structure can be partioned into four quadrants, each representing a different structural component of the HGP code. The nodes in the upper-left and lower-right blocks denote qubits. The nodes in the upper-right block denote $Z$-stabilizer generators; these correspond to the rows of $H_Z$. Similarly, the nodes in the lower-left block denote $X$-stabilizer generators; these correspond to the rows of $H_X$.}
    \label{fig:HGP_TannerGraph}
\end{figure}

HGP codes have a geometrically rich Tanner graph structure which can be visualized as the cartesian product of the Tanner graphs for the two input classical codes as shown in Fig.~\ref{fig:HGP_TannerGraph}. The subgraph corresponding to each row and column in this Tanner graph block structure can be understood as the classical Tanner graph for one of the classical codes used in the construction. As shown in the figure, qubits are represented by circular nodes, and stabilizer checks of both types are represented by square nodes. Additional details regarding this construction are discussed in Appendix~\ref{app:HGP_construction}.

Surface codes may also be recovered as a special case of hypergraph product code. Using parity check matrices $H_1$ and $H_2$ for a classical repetition code, $\text{HGP}(H_1,H_2)$ is exactly the toric surface code. Hence, adapting the multiplexing strategies discussed in Sec.~\ref{sec_SC} to this more general class of codes is a natural next question. However, the linear-time maximum-likelihood generalization of the peeling decoder~\cite{delfosse2020linear} used in our previous simulations is only defined for the special case of the surface code. This decoder leverages the lattice structure of the surface code to ensure that the erasure subgraph can be completely peeled (Appendix~\ref{app:peeling_surface}), but this technique does not apply to generic HGP codes. Instead, we introduce another generalization of the peeling algorithm to extend our numerical analysis to HGP codes as well.

\subsection{Pruned-Peeling + VH Decoder}
\label{sec_HGP_decoder}
The \textit{pruned peeling + VH decoder}~\cite{connolly2024fast} is a generalization of the peeling algorithm (outlined in Appendix~\ref{app:peeling}) specifically designed for HGP codes, where $VH$ stands for Vertical-Horizontal. It has quadratic complexity and close to maximum-likelihood performance at low erasure rate, making it practically useful for our simulations. This decoder is a modified version of the standard classical peeling decoder based on analysis and correction of two common types of \textit{stopping sets}, which are patterns of erased qubits that cannot be corrected by simple peeling.

A \textit{stabilizer stopping set} occurs when the erasure pattern covers the qubit support of an $X$- or $Z$-type stabilizer. The \textit{pruned peeling decoder} attempts to fix these by removing a qubit from the erasure, thus "breaking" the stabilizer support and possibly allowing the peeling algorithm to become unstuck. \textit{Classical stopping sets} are patterns of erased qubits supported entirely on a single row or column in the HGP Tanner graph block structure of Fig.~\ref{fig:HGP_TannerGraph}; any stopping set for a HGP code can be decomposed into a union of components of this form. The \textit{VH decoder} algorithm attempts to order and efficiently solve each of these classical stopping sets in sequence. The name ``vertical-horizontal'' refers to the fact that, in the HGP Tanner graph picture, individual classical stopping sets are either entirely supported on a column (vertical) or a row (horizontal). The combination of these decoding strategies is referred to as the \textit{combined decoder} (peeling + pruned peeling + VH). A more detailed explanation is included in Appendix~\ref{app:peeling_HGP}.

The combined decoder is not a maximum likelihood decoder. In addition to logical errors, there still exist patterns of erased qubits where the decoder becomes stuck in a stopping set, leading to a decoder failure. We introduce the term \textit{error recovery failure} to refer to both decoder failures and logical errors (discussed in Appendix~\ref{app:HGP_decoder}). Our numerical simulations for HGP codes are always with respect to the \textit{error recovery failure rate}.

\begin{figure}
    \centering
    \includegraphics[width = 1 \columnwidth]{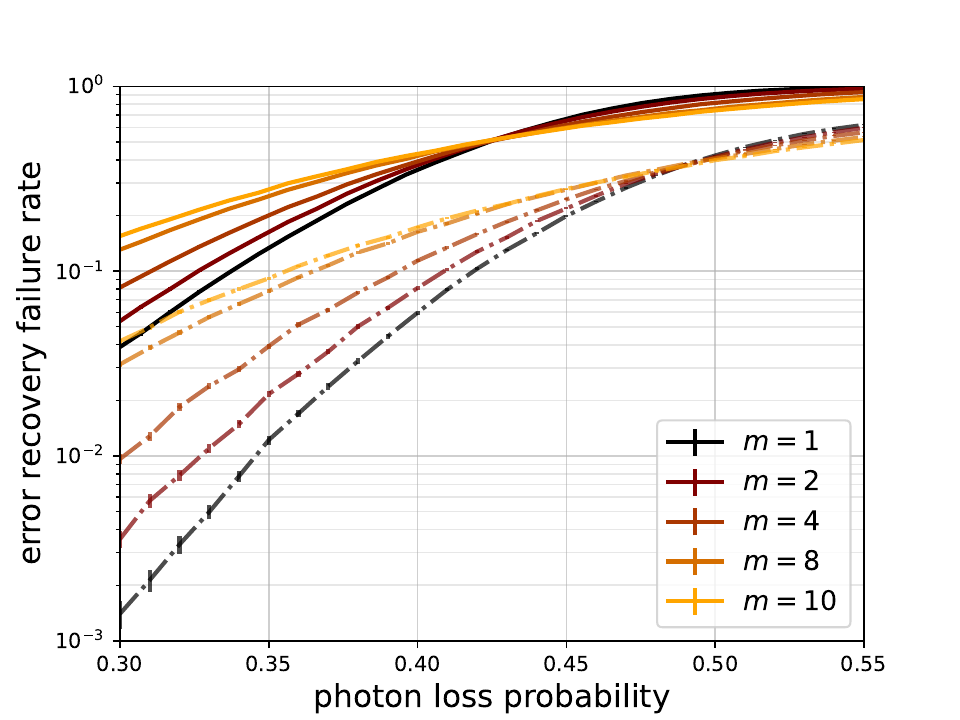}
    \caption{Performance of non-ML \textit{combined decoder} (peeling + pruned peeling + VH) as shown in solid curves and  the \textit{surface code peeling decoder} as shown in dashed curves applied to the $[\![200,2,10]\!]$ surface code using the uniformly random assignment strategy with different numbers of qubits in a single photon, $m$.
    }
    \label{fig:Toric10_VH_performance}
\end{figure}

Fig.~\ref{fig:Toric10_VH_performance} shows the numerical performance of the combined decoder applied to the $10\times10$ surface code for comparison with the replotted data from Fig.~\ref{fig:m}, which shows the performance of the ML decoder applied to this same code. The performance degradation is explained by the presence of decoder failures that do not exist in the ML case. Even though it is not ML, the pruned peeling + VH decoder is still practically useful for our numerical simulations since decoder failures are infrequent at low erasure rates. 

\subsection{Assignment Strategies for HGP Codes}
\label{sec_HGP_assignment}
As with the surface code, quantum multiplexing can also be utilized with HGP codes. In this section, we analyze the performance of HGP code communication in scenario \textbf{(C)}. The scenarios previously proposed in Sec.~\ref{sec_SC} are also valid for HGP codes, but unlike the special case of the surface code, the distance between any two qubits in a generic HGP code is not easily inferred from a grid. Hence, we do not consider the previously introduced strategies which use distance. We also introduce several new strategies for HGP codes based on stopping sets for the pruned peeling + VH decoder. These strategies are summarized in Table~\ref{table:HGP_assignment_strategies}, and technical details are included in Appendix~\ref{app:HGP_strategies}.

\begin{table*}[htbp]
\scalebox{0.85}[0.85]{
\begin{tabular}{ccccc}

\begin{tikzpicture}[baseline]
\node (nw) [inner sep=0] at (-5/2,5/2) {};
\node (n) [inner sep=0] at (0,5/2) {};
\node (ne) [inner sep=0] at (3/2,5/2) {};
\node (w) [inner sep=0] at (-5/2,0) {};
\node (c) [inner sep=0] at (0,0) {};
\node (e) [inner sep=0] at (3/2,0) {};
\node (sw) [inner sep=0] at (-5/2,-3/2) {};
\node (s) [inner sep=0] at (0,-3/2) {};
\node (se) [inner sep=0] at (3/2,-3/2) {};

\path (nw) edge[line width=1,opacity=1] (ne);
\path (w) edge[line width=1,dashed,opacity=1] (e);
\path (sw) edge[line width=1,opacity=1] (se);
\path (nw) edge[line width=1,opacity=1] (sw);
\path (n) edge[line width=1,dashed,opacity=1] (s);
\path (ne) edge[line width=1,opacity=1] (se);

\node (-44) [circle,draw,inner sep=3pt,line width=1,opacity=1,fill=Apricot] at (-4/2,4/2) {};
\node (-34) [circle,draw,inner sep=3pt,line width=1,opacity=1,fill=CornflowerBlue] at (-3/2,4/2) {};
\node (-24) [circle,draw,inner sep=3pt,line width=1,opacity=1,fill=GreenYellow] at (-2/2,4/2) {};
\node (-14) [circle,draw,inner sep=3pt,line width=1,opacity=1,fill=Orchid] at (-1/2,4/2) {};
\node (-43) [circle,draw,inner sep=3pt,line width=1,opacity=1,fill=Apricot] at (-4/2,3/2) {};
\node (-33) [circle,draw,inner sep=3pt,line width=1,opacity=1,fill=CornflowerBlue] at (-3/2,3/2) {};
\node (-23) [circle,draw,inner sep=3pt,line width=1,opacity=1,fill=GreenYellow] at (-2/2,3/2) {};
\node (-13) [circle,draw,inner sep=3pt,line width=1,opacity=1,fill=Orchid] at (-1/2,3/2) {};
\node (-42) [circle,draw,inner sep=3pt,line width=1,opacity=1,fill=Apricot] at (-4/2,2/2) {};
\node (-32) [circle,draw,inner sep=3pt,line width=1,opacity=1,fill=CornflowerBlue] at (-3/2,2/2) {};
\node (-22) [circle,draw,inner sep=3pt,line width=1,opacity=1,fill=GreenYellow] at (-2/2,2/2) {};
\node (-12) [circle,draw,inner sep=3pt,line width=1,opacity=1,fill=Orchid] at (-1/2,2/2) {};
\node (-41) [circle,draw,inner sep=3pt,line width=1,opacity=1,fill=Apricot] at (-4/2,1/2) {};
\node (-31) [circle,draw,inner sep=3pt,line width=1,opacity=1,fill=CornflowerBlue] at (-3/2,1/2) {};
\node (-21) [circle,draw,inner sep=3pt,line width=1,opacity=1,fill=GreenYellow] at (-2/2,1/2) {};
\node (-11) [circle,draw,inner sep=3pt,line width=1,opacity=1,fill=Orchid] at (-1/2,1/2) {};

\node (14) [rectangle,draw,inner sep=4pt,line width=1,opacity=1] at (1/2,4/2) {};
\node (24) [rectangle,draw,inner sep=4pt,line width=1,opacity=1] at (2/2,4/2) {};
\node (13) [rectangle,draw,inner sep=4pt,line width=1,opacity=1] at (1/2,3/2) {};
\node (23) [rectangle,draw,inner sep=4pt,line width=1,opacity=1] at (2/2,3/2) {};
\node (12) [rectangle,draw,inner sep=4pt,line width=1,opacity=1] at (1/2,2/2) {};
\node (22) [rectangle,draw,inner sep=4pt,line width=1,opacity=1] at (2/2,2/2) {};
\node (11) [rectangle,draw,inner sep=4pt,line width=1,opacity=1] at (1/2,1/2) {};
\node (21) [rectangle,draw,inner sep=4pt,line width=1,opacity=1] at (2/2,1/2) {};

\node (-4-1) [rectangle,draw,inner sep=4pt,line width=1,opacity=1,fill=Apricot] at (-4/2,-1/2) {};
\node (-3-1) [rectangle,draw,inner sep=4pt,line width=1,opacity=1] at (-3/2,-1/2) {};
\node (-2-1) [rectangle,draw,inner sep=4pt,line width=1,opacity=1,fill=GreenYellow] at (-2/2,-1/2) {};
\node (-1-1) [rectangle,draw,inner sep=4pt,line width=1,opacity=1] at (-1/2,-1/2) {};
\node (-4-2) [rectangle,draw,inner sep=4pt,line width=1,opacity=1] at (-4/2,-2/2) {};
\node (-3-2) [rectangle,draw,inner sep=4pt,line width=1,opacity=1,fill=CornflowerBlue] at (-3/2,-2/2) {};
\node (-2-2) [rectangle,draw,inner sep=4pt,line width=1,opacity=1] at (-2/2,-2/2) {};
\node (-1-2) [rectangle,draw,inner sep=4pt,line width=1,opacity=1,fill=Orchid] at (-1/2,-2/2) {};

\node (1-1) [circle,draw,inner sep=3pt,line width=1,opacity=1,fill=Apricot] at (1/2,-1/2) {};
\node (2-1) [circle,draw,inner sep=3pt,line width=1,opacity=1,fill=GreenYellow] at (2/2,-1/2) {};
\node (1-2) [circle,draw,inner sep=3pt,line width=1,opacity=1,fill=CornflowerBlue] at (1/2,-2/2) {};
\node (2-2) [circle,draw,inner sep=3pt,line width=1,opacity=1,fill=Orchid] at (2/2,-2/2) {};

\path (-34) edge[line width=0.75,bend left=-40,looseness=0.75,color=CornflowerBlue] (-3-2);
\path (-33) edge[line width=0.75,bend left=-40,looseness=0.75,color=CornflowerBlue] (-3-2);
\path (-32) edge[line width=0.75,bend left=-40,looseness=0.75,color=CornflowerBlue] (-3-2);
\path (-31) edge[line width=0.75,bend left=-40,looseness=0.75,color=CornflowerBlue] (-3-2);
\path (-3-2) edge[line width=0.75,bend left=40,looseness=0.75,color=CornflowerBlue] (1-2);

\end{tikzpicture}

 & 

\begin{tikzpicture}[baseline]
\node (nw) [inner sep=0] at (-5/2,5/2) {};
\node (n) [inner sep=0] at (0,5/2) {};
\node (ne) [inner sep=0] at (3/2,5/2) {};
\node (w) [inner sep=0] at (-5/2,0) {};
\node (c) [inner sep=0] at (0,0) {};
\node (e) [inner sep=0] at (3/2,0) {};
\node (sw) [inner sep=0] at (-5/2,-3/2) {};
\node (s) [inner sep=0] at (0,-3/2) {};
\node (se) [inner sep=0] at (3/2,-3/2) {};

\path (nw) edge[line width=1,opacity=1] (ne);
\path (w) edge[line width=1,dashed,opacity=1] (e);
\path (sw) edge[line width=1,opacity=1] (se);
\path (nw) edge[line width=1,opacity=1] (sw);
\path (n) edge[line width=1,dashed,opacity=1] (s);
\path (ne) edge[line width=1,opacity=1] (se);

\node (-44) [circle,draw,inner sep=3pt,line width=1,opacity=1,fill=Apricot] at (-4/2,4/2) {};
\node (-34) [circle,draw,inner sep=3pt,line width=1,opacity=1,fill=GreenYellow] at (-3/2,4/2) {};
\node (-24) [circle,draw,inner sep=3pt,line width=1,opacity=1,fill=CornflowerBlue] at (-2/2,4/2) {};
\node (-14) [circle,draw,inner sep=3pt,line width=1,opacity=1,fill=Orchid] at (-1/2,4/2) {};
\node (-43) [circle,draw,inner sep=3pt,line width=1,opacity=1,fill=CornflowerBlue] at (-4/2,3/2) {};
\node (-33) [circle,draw,inner sep=3pt,line width=1,opacity=1,fill=Maroon] at (-3/2,3/2) {};
\node (-23) [circle,draw,inner sep=3pt,line width=1,opacity=1,fill=Orchid] at (-2/2,3/2) {};
\node (-13) [circle,draw,inner sep=3pt,line width=1,opacity=1,fill=GreenYellow] at (-1/2,3/2) {};
\node (-42) [circle,draw,inner sep=3pt,line width=1,opacity=1,fill=Orchid] at (-4/2,2/2) {};
\node (-32) [circle,draw,inner sep=3pt,line width=1,opacity=1,fill=CornflowerBlue] at (-3/2,2/2) {};
\node (-22) [circle,draw,inner sep=3pt,line width=1,opacity=1,fill=Apricot] at (-2/2,2/2) {};
\node (-12) [circle,draw,inner sep=3pt,line width=1,opacity=1,fill=Maroon] at (-1/2,2/2) {};
\node (-41) [circle,draw,inner sep=3pt,line width=1,opacity=1,fill=Maroon] at (-4/2,1/2) {};
\node (-31) [circle,draw,inner sep=3pt,line width=1,opacity=1,fill=Apricot] at (-3/2,1/2) {};
\node (-21) [circle,draw,inner sep=3pt,line width=1,opacity=1,fill=GreenYellow] at (-2/2,1/2) {};
\node (-11) [circle,draw,inner sep=3pt,line width=1,opacity=1,fill=CornflowerBlue] at (-1/2,1/2) {};

\node (14) [rectangle,draw,inner sep=4pt,line width=1,opacity=1] at (1/2,4/2) {};
\node (24) [rectangle,draw,inner sep=4pt,line width=1,opacity=1] at (2/2,4/2) {};
\node (13) [rectangle,draw,inner sep=4pt,line width=1,opacity=1] at (1/2,3/2) {};
\node (23) [rectangle,draw,inner sep=4pt,line width=1,opacity=1] at (2/2,3/2) {};
\node (12) [rectangle,draw,inner sep=4pt,line width=1,opacity=1] at (1/2,2/2) {};
\node (22) [rectangle,draw,inner sep=4pt,line width=1,opacity=1] at (2/2,2/2) {};
\node (11) [rectangle,draw,inner sep=4pt,line width=1,opacity=1] at (1/2,1/2) {};
\node (21) [rectangle,draw,inner sep=4pt,line width=1,opacity=1] at (2/2,1/2) {};

\node (-4-1) [rectangle,draw,inner sep=4pt,line width=1,opacity=1] at (-4/2,-1/2) {};
\node (-3-1) [rectangle,draw,inner sep=4pt,line width=1,opacity=1] at (-3/2,-1/2) {};
\node (-2-1) [rectangle,draw,inner sep=4pt,line width=1,opacity=1] at (-2/2,-1/2) {};
\node (-1-1) [rectangle,draw,inner sep=4pt,line width=1,opacity=1] at (-1/2,-1/2) {};
\node (-4-2) [rectangle,draw,inner sep=4pt,line width=1,opacity=1] at (-4/2,-2/2) {};
\node (-3-2) [rectangle,draw,inner sep=4pt,line width=1,opacity=1] at (-3/2,-2/2) {};
\node (-2-2) [rectangle,draw,inner sep=4pt,line width=1,opacity=1] at (-2/2,-2/2) {};
\node (-1-2) [rectangle,draw,inner sep=4pt,line width=1,opacity=1] at (-1/2,-2/2) {};

\node (1-1) [circle,draw,inner sep=3pt,line width=1,opacity=1,fill=Maroon] at (1/2,-1/2) {};
\node (2-1) [circle,draw,inner sep=3pt,line width=1,opacity=1,fill=Apricot] at (2/2,-1/2) {};
\node (1-2) [circle,draw,inner sep=3pt,line width=1,opacity=1,fill=Orchid] at (1/2,-2/2) {};
\node (2-2) [circle,draw,inner sep=3pt,line width=1,opacity=1,fill=GreenYellow] at (2/2,-2/2) {};

\end{tikzpicture}

 &

\begin{tikzpicture}[baseline]
\node (nw) [inner sep=0] at (-5/2,5/2) {};
\node (n) [inner sep=0] at (0,5/2) {};
\node (ne) [inner sep=0] at (3/2,5/2) {};
\node (w) [inner sep=0] at (-5/2,0) {};
\node (c) [inner sep=0] at (0,0) {};
\node (e) [inner sep=0] at (3/2,0) {};
\node (sw) [inner sep=0] at (-5/2,-3/2) {};
\node (s) [inner sep=0] at (0,-3/2) {};
\node (se) [inner sep=0] at (3/2,-3/2) {};

\path (nw) edge[line width=1,opacity=1] (ne);
\path (w) edge[line width=1,dashed,opacity=1] (e);
\path (sw) edge[line width=1,opacity=1] (se);
\path (nw) edge[line width=1,opacity=1] (sw);
\path (n) edge[line width=1,dashed,opacity=1] (s);
\path (ne) edge[line width=1,opacity=1] (se);

\node (-44) [circle,draw,inner sep=3pt,line width=1,opacity=1,fill=Apricot] at (-4/2,4/2) {};
\node (-34) [circle,draw,inner sep=3pt,line width=1,opacity=1,fill=Apricot] at (-3/2,4/2) {};
\node (-24) [circle,draw,inner sep=3pt,line width=1,opacity=1,fill=Maroon] at (-2/2,4/2) {};
\node (-14) [circle,draw,inner sep=3pt,line width=1,opacity=1,fill=Maroon] at (-1/2,4/2) {};
\node (-43) [circle,draw,inner sep=3pt,line width=1,opacity=1,fill=CornflowerBlue] at (-4/2,3/2) {};
\node (-33) [circle,draw,inner sep=3pt,line width=1,opacity=1,fill=CornflowerBlue] at (-3/2,3/2) {};
\node (-23) [circle,draw,inner sep=3pt,line width=1,opacity=1,fill=Grayyellow] at (-2/2,3/2) {};
\node (-13) [circle,draw,inner sep=3pt,line width=1,opacity=1,fill=Grayyellow] at (-1/2,3/2) {};
\node (-42) [circle,draw,inner sep=3pt,line width=1,opacity=1,fill=GreenYellow] at (-4/2,2/2) {};
\node (-32) [circle,draw,inner sep=3pt,line width=1,opacity=1,fill=GreenYellow] at (-3/2,2/2) {};
\node (-22) [circle,draw,inner sep=3pt,line width=1,opacity=1,fill=Grayred] at (-2/2,2/2) {};
\node (-12) [circle,draw,inner sep=3pt,line width=1,opacity=1,fill=Grayred] at (-1/2,2/2) {};
\node (-41) [circle,draw,inner sep=3pt,line width=1,opacity=1,fill=Orchid] at (-4/2,1/2) {};
\node (-31) [circle,draw,inner sep=3pt,line width=1,opacity=1,fill=Orchid] at (-3/2,1/2) {};
\node (-21) [circle,draw,inner sep=3pt,line width=1,opacity=1,fill=Tan] at (-2/2,1/2) {};
\node (-11) [circle,draw,inner sep=3pt,line width=1,opacity=1,fill=Tan] at (-1/2,1/2) {};

\node (14) [rectangle,draw,inner sep=4pt,line width=1,opacity=1] at (1/2,4/2) {};
\node (24) [rectangle,draw,inner sep=4pt,line width=1,opacity=1] at (2/2,4/2) {};
\node (13) [rectangle,draw,inner sep=4pt,line width=1,opacity=1] at (1/2,3/2) {};
\node (23) [rectangle,draw,inner sep=4pt,line width=1,opacity=1] at (2/2,3/2) {};
\node (12) [rectangle,draw,inner sep=4pt,line width=1,opacity=1] at (1/2,2/2) {};
\node (22) [rectangle,draw,inner sep=4pt,line width=1,opacity=1] at (2/2,2/2) {};
\node (11) [rectangle,draw,inner sep=4pt,line width=1,opacity=1] at (1/2,1/2) {};
\node (21) [rectangle,draw,inner sep=4pt,line width=1,opacity=1] at (2/2,1/2) {};

\node (-4-1) [rectangle,draw,inner sep=4pt,line width=1,opacity=1] at (-4/2,-1/2) {};
\node (-3-1) [rectangle,draw,inner sep=4pt,line width=1,opacity=1] at (-3/2,-1/2) {};
\node (-2-1) [rectangle,draw,inner sep=4pt,line width=1,opacity=1] at (-2/2,-1/2) {};
\node (-1-1) [rectangle,draw,inner sep=4pt,line width=1,opacity=1] at (-1/2,-1/2) {};
\node (-4-2) [rectangle,draw,inner sep=4pt,line width=1,opacity=1] at (-4/2,-2/2) {};
\node (-3-2) [rectangle,draw,inner sep=4pt,line width=1,opacity=1] at (-3/2,-2/2) {};
\node (-2-2) [rectangle,draw,inner sep=4pt,line width=1,opacity=1] at (-2/2,-2/2) {};
\node (-1-2) [rectangle,draw,inner sep=4pt,line width=1,opacity=1] at (-1/2,-2/2) {};

\node (1-1) [circle,draw,inner sep=3pt,line width=1,opacity=1,fill=Lavender] at (1/2,-1/2) {};
\node (2-1) [circle,draw,inner sep=3pt,line width=1,opacity=1,fill=LimeGreen] at (2/2,-1/2) {};
\node (1-2) [circle,draw,inner sep=3pt,line width=1,opacity=1,fill=Lavender] at (1/2,-2/2) {};
\node (2-2) [circle,draw,inner sep=3pt,line width=1,opacity=1,fill=LimeGreen] at (2/2,-2/2) {};

\node (-4p54) [inner sep=0] at (-4.75/2,4/2) {};
\node (-p54) [inner sep=0] at (-0.25/2,4/2) {};
\path (-4p54) edge[line width=0.75,->] (-p54);

\node (-4p53) [inner sep=0] at (-4.75/2,3/2) {};
\node (-p53) [inner sep=0] at (-0.25/2,3/2) {};
\path (-4p53) edge[line width=0.75,->] (-p53);

\node (-4p52) [inner sep=0] at (-4.75/2,2/2) {};
\node (-p52) [inner sep=0] at (-0.25/2,2/2) {};
\path (-4p52) edge[line width=0.75,->] (-p52);

\node (-4p51) [inner sep=0] at (-4.75/2,1/2) {};
\node (-p51) [inner sep=0] at (-0.25/2,1/2) {};
\path (-4p51) edge[line width=0.75,->] (-p51);

\node (1-p5) [inner sep=0] at (1/2,-0.25/2) {};
\node (1-2p5) [inner sep=0] at (1/2,-2.75/2) {};
\path (1-p5) edge[line width=0.75,->] (1-2p5);

\node (2-p5) [inner sep=0] at (2/2,-0.25/2) {};
\node (2-2p5) [inner sep=0] at (2/2,-2.75/2) {};
\path (2-p5) edge[line width=0.75,->] (2-2p5);

\end{tikzpicture}

 &

\begin{tikzpicture}[baseline]
\node (nw) [inner sep=0] at (-5/2,5/2) {};
\node (n) [inner sep=0] at (0,5/2) {};
\node (ne) [inner sep=0] at (3/2,5/2) {};
\node (w) [inner sep=0] at (-5/2,0) {};
\node (c) [inner sep=0] at (0,0) {};
\node (e) [inner sep=0] at (3/2,0) {};
\node (sw) [inner sep=0] at (-5/2,-3/2) {};
\node (s) [inner sep=0] at (0,-3/2) {};
\node (se) [inner sep=0] at (3/2,-3/2) {};

\path (nw) edge[line width=1,opacity=1] (ne);
\path (w) edge[line width=1,dashed,opacity=1] (e);
\path (sw) edge[line width=1,opacity=1] (se);
\path (nw) edge[line width=1,opacity=1] (sw);
\path (n) edge[line width=1,dashed,opacity=1] (s);
\path (ne) edge[line width=1,opacity=1] (se);

\node (-44) [circle,draw,inner sep=3pt,line width=1,opacity=1,fill=Grayyellow] at (-4/2,4/2) {};
\node (-34) [circle,draw,inner sep=3pt,line width=1,opacity=1,fill=Grayred] at (-3/2,4/2) {};
\node (-24) [circle,draw,inner sep=3pt,line width=1,opacity=1,fill=Tan] at (-2/2,4/2) {};
\node (-14) [circle,draw,inner sep=3pt,line width=1,opacity=1,fill=Maroon] at (-1/2,4/2) {};
\node (-43) [circle,draw,inner sep=3pt,line width=1,opacity=1,fill=Maroon] at (-4/2,3/2) {};
\node (-33) [circle,draw,inner sep=3pt,line width=1,opacity=1,fill=Grayyellow] at (-3/2,3/2) {};
\node (-23) [circle,draw,inner sep=3pt,line width=1,opacity=1,fill=Grayred] at (-2/2,3/2) {};
\node (-13) [circle,draw,inner sep=3pt,line width=1,opacity=1,fill=Tan] at (-1/2,3/2) {};
\node (-42) [circle,draw,inner sep=3pt,line width=1,opacity=1,fill=Orchid] at (-4/2,2/2) {};
\node (-32) [circle,draw,inner sep=3pt,line width=1,opacity=1,fill=Apricot] at (-3/2,2/2) {};
\node (-22) [circle,draw,inner sep=3pt,line width=1,opacity=1,fill=CornflowerBlue] at (-2/2,2/2) {};
\node (-12) [circle,draw,inner sep=3pt,line width=1,opacity=1,fill=GreenYellow] at (-1/2,2/2) {};
\node (-41) [circle,draw,inner sep=3pt,line width=1,opacity=1,fill=GreenYellow] at (-4/2,1/2) {};
\node (-31) [circle,draw,inner sep=3pt,line width=1,opacity=1,fill=Orchid] at (-3/2,1/2) {};
\node (-21) [circle,draw,inner sep=3pt,line width=1,opacity=1,fill=Apricot] at (-2/2,1/2) {};
\node (-11) [circle,draw,inner sep=3pt,line width=1,opacity=1,fill=CornflowerBlue] at (-1/2,1/2) {};

\node (14) [rectangle,draw,inner sep=4pt,line width=1,opacity=1] at (1/2,4/2) {};
\node (24) [rectangle,draw,inner sep=4pt,line width=1,opacity=1] at (2/2,4/2) {};
\node (13) [rectangle,draw,inner sep=4pt,line width=1,opacity=1] at (1/2,3/2) {};
\node (23) [rectangle,draw,inner sep=4pt,line width=1,opacity=1] at (2/2,3/2) {};
\node (12) [rectangle,draw,inner sep=4pt,line width=1,opacity=1] at (1/2,2/2) {};
\node (22) [rectangle,draw,inner sep=4pt,line width=1,opacity=1] at (2/2,2/2) {};
\node (11) [rectangle,draw,inner sep=4pt,line width=1,opacity=1] at (1/2,1/2) {};
\node (21) [rectangle,draw,inner sep=4pt,line width=1,opacity=1] at (2/2,1/2) {};

\node (-4-1) [rectangle,draw,inner sep=4pt,line width=1,opacity=1] at (-4/2,-1/2) {};
\node (-3-1) [rectangle,draw,inner sep=4pt,line width=1,opacity=1] at (-3/2,-1/2) {};
\node (-2-1) [rectangle,draw,inner sep=4pt,line width=1,opacity=1] at (-2/2,-1/2) {};
\node (-1-1) [rectangle,draw,inner sep=4pt,line width=1,opacity=1] at (-1/2,-1/2) {};
\node (-4-2) [rectangle,draw,inner sep=4pt,line width=1,opacity=1] at (-4/2,-2/2) {};
\node (-3-2) [rectangle,draw,inner sep=4pt,line width=1,opacity=1] at (-3/2,-2/2) {};
\node (-2-2) [rectangle,draw,inner sep=4pt,line width=1,opacity=1] at (-2/2,-2/2) {};
\node (-1-2) [rectangle,draw,inner sep=4pt,line width=1,opacity=1] at (-1/2,-2/2) {};

\node (1-1) [circle,draw,inner sep=3pt,line width=1,opacity=1,fill=LimeGreen] at (1/2,-1/2) {};
\node (2-1) [circle,draw,inner sep=3pt,line width=1,opacity=1,fill=Lavender] at (2/2,-1/2) {};
\node (1-2) [circle,draw,inner sep=3pt,line width=1,opacity=1,fill=Lavender] at (1/2,-2/2) {};
\node (2-2) [circle,draw,inner sep=3pt,line width=1,opacity=1,fill=LimeGreen] at (2/2,-2/2) {};

\node (-4p54p5) [inner sep=0] at (-4.5/2,4.5/2) {};
\node (-p5p5) [inner sep=0] at (-0.5/2,0.5/2) {};
\path (-4p54p5) edge[line width=0.75,->] (-p5p5);

\node (-4p53p5) [inner sep=0] at (-4.5/2,3.5/2) {};
\node (-1p5p5) [inner sep=0] at (-1.5/2,0.5/2) {};
\path (-4p53p5) edge[line width=0.75,->] (-1p5p5);

\node (-4p52p5) [inner sep=0] at (-4.5/2,2.5/2) {};
\node (-2p5p5) [inner sep=0] at (-2.5/2,0.5/2) {};
\path (-4p52p5) edge[line width=0.75,->] (-2p5p5);

\node (-4p51p5) [inner sep=0] at (-4.5/2,1.5/2) {};
\node (-3p5p5) [inner sep=0] at (-3.5/2,0.5/2) {};
\path (-4p51p5) edge[line width=0.75,->] (-3p5p5);

\node (-3p54p5) [inner sep=0] at (-3.5/2,4.5/2) {};
\node (-p51p5) [inner sep=0] at (-0.5/2,1.5/2) {};
\path (-3p54p5) edge[line width=0.75,->] (-p51p5);

\node (-2p54p5) [inner sep=0] at (-2.5/2,4.5/2) {};
\node (-p52p5) [inner sep=0] at (-0.5/2,2.5/2) {};
\path (-2p54p5) edge[line width=0.75,->] (-p52p5);

\node (-1p54p5) [inner sep=0] at (-1.5/2,4.5/2) {};
\node (-p53p5) [inner sep=0] at (-0.5/2,3.5/2) {};
\path (-1p54p5) edge[line width=0.75,->] (-p53p5);

\node (p5-p5) [inner sep=0] at (0.5/2,-0.5/2) {};
\node (2p5-2p5) [inner sep=0] at (2.5/2,-2.5/2) {};
\path (p5-p5) edge[line width=0.75,->] (2p5-2p5);

\node (p5-1p5) [inner sep=0] at (0.5/2,-1.5/2) {};
\node (1p5-2p5) [inner sep=0] at (1.5/2,-2.5/2) {};
\path (p5-1p5) edge[line width=0.75,->] (1p5-2p5);

\node (1p5-p5) [inner sep=0] at (1.5/2,-0.5/2) {};
\node (2p5-1p5) [inner sep=0] at (2.5/2,-1.5/2) {};
\path (1p5-p5) edge[line width=0.75,->] (2p5-1p5);

\end{tikzpicture}

 &
 \\ 
 \\
 \cline{1-4}
 {\bf Strategy ii. Stabilizer} & {\bf Strategy iii. Sudoku} & {\bf Strategy iv. Row-Column} & {\bf Strategy v. Diagonal} \\ 
 \cline{1-4}
\end{tabular}}
\caption{Examples of four different photon assignment strategies for the simple HGP code shown in Fig.~\ref{fig:HGP_TannerGraph}. {\bf (ii.)} Each photon in the \textit{stabilizer strategy} is the qubit-support of an $X$ or $Z$-type stabilizer generator, identified as a row of $H_X$ or $H_Z$. The number of qubits per photon is a fraction or multiple of the weight of the corresponding row. {\bf (iii.)} In the \textit{sudoku strategy}, each qubit of a given photon is contained in a different row or column of the HGP Tanner graph. {\bf (iv.)} Using the \textit{row-column strategy}, each qubit of a given photon is contained in the same row or column of the HGP Tanner graph. {\bf (v.)} Photons from the \textit{diagonal strategy} contain qubits from the same diagonal slice of the HGP Tanner graph, allowing diagonal lines to wrap around. For strategies iii., iv., and v., the number of qubits per photon is a fraction or multiple of the shortest side length in the block structure.}
\label{table:HGP_assignment_strategies}
\end{table*}

{\bf Strategy i: random}
The simplest assignment strategy is based on assigning qubits to photons uniformly at random.

{\bf Strategy ii: stabilizer} The stabilizer assignment strategy assigns qubits to photons so that photons correspond to the qubit-support of a stabilizer. This strategy is motivated by the fact that the pruned peeling decoder is designed to correct erased stabilizers.

{\bf Strategy iii: sudoku}
The sudoku strategy assigns qubits to photons at random subject to the condition that qubits within a given photon come from different rows and columns in the HGP Tanner graph structure. It is motivated by the goal of reducing classical stopping sets, a common source of peeling decoder failures for HGP codes. We name this the sudoku strategy due to its resemblance to the popular game.

{\bf Strategy iv: row-column}
In contrast to sudoku, the row-column strategy chooses qubits in a given photon from the same row or column of the HGP Tanner graph structure. It seeks to maximize the number of classical stopping sets and hence decoder failures. The row-column strategy can be interpreted as a worst-case scenario.

{\bf Strategy v: diagonal}
The diagonal assignment strategy is based on dividing the qubit blocks in the HGP Tanner graph into diagonal slices. Qubits from the same diagonal slice are assigned to the same photon. This is a modified version of the sudoku strategy which does not use randomness but still seeks to minimize classical stopping sets and hence decoder failures.

\begin{figure}
    \centering
    \includegraphics[width = 1 \columnwidth]{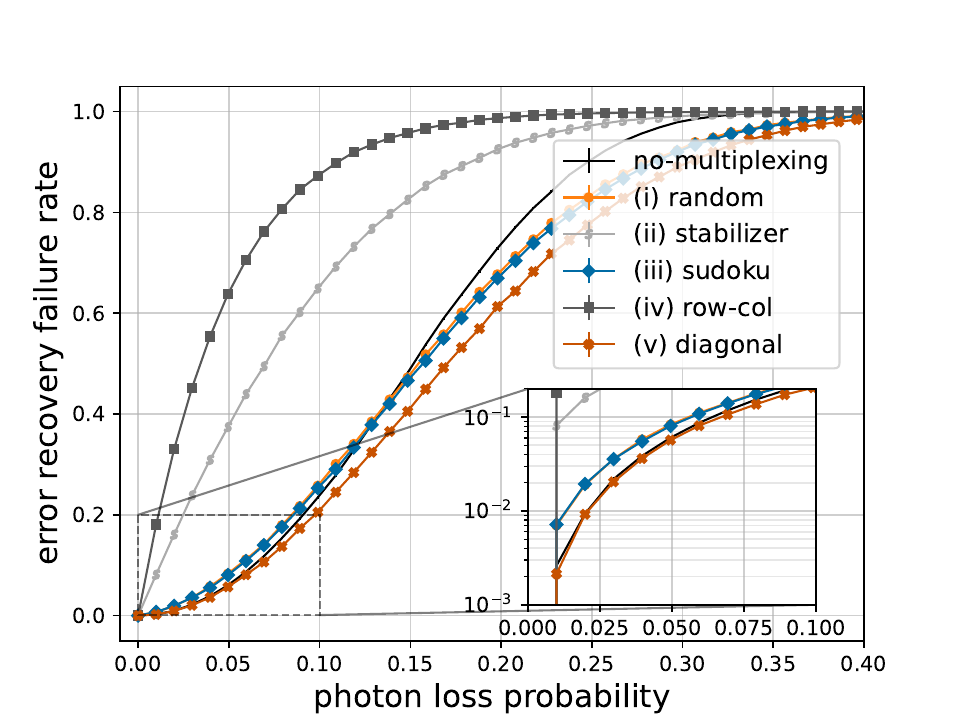}
    \caption{Multiplexing decoder performance for a [\![320,82]\!] non-equal block ($16\times 16$ and $8\times 8$) HGP code at fixed $m=8$. In this example, strategy (v) diagonal outperforms all other strategies, including the no-multiplexing case.}
    \label{fig:n320_m8}
\end{figure}

\begin{figure*}
\begin{tabular}{cc}
     \begin{minipage}[b]{0.5\linewidth}
     \centering
     \includegraphics[width = 1 \columnwidth]{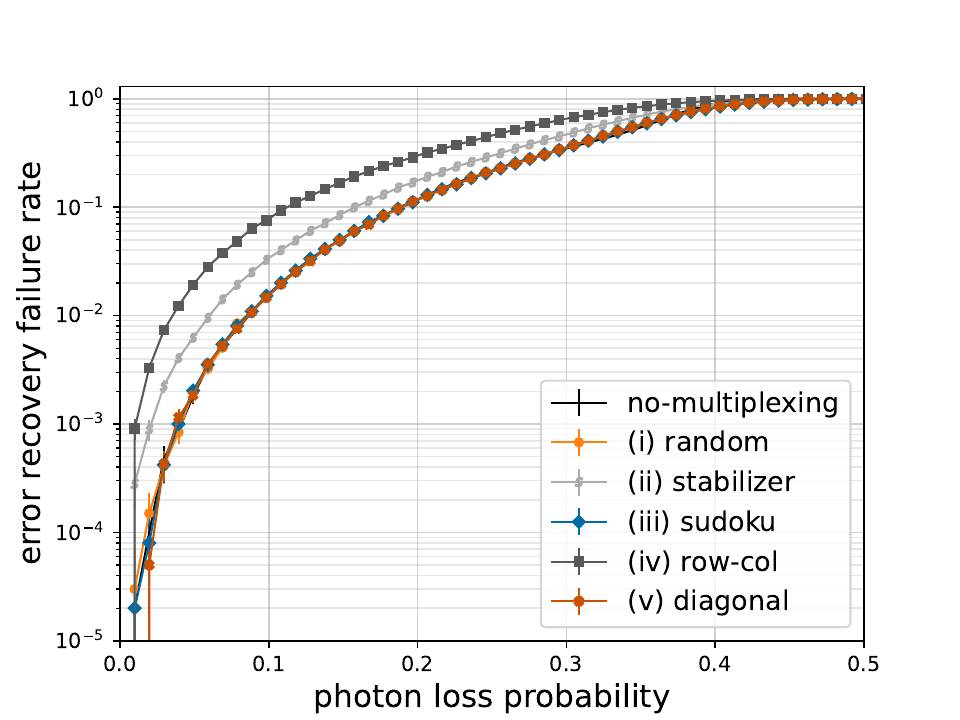}
     \end{minipage}
    &
     \begin{minipage}[b]{0.5\linewidth}
     \centering
     \includegraphics[width = 1 \columnwidth]{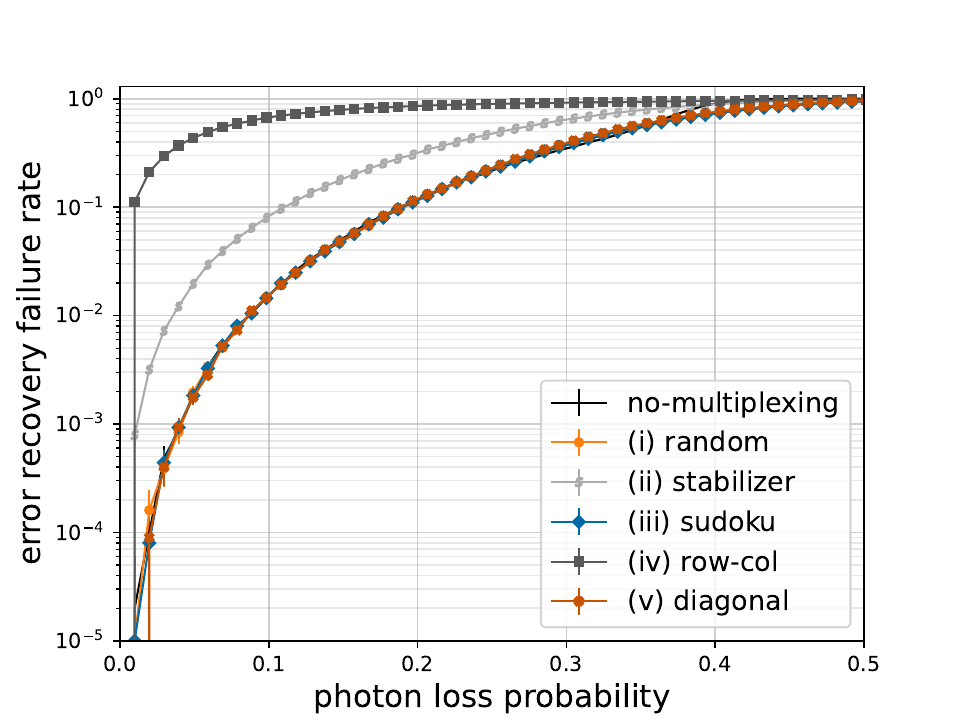}
     \end{minipage}
    \\
 (A) {\bf $m=4$} & (B) {\bf $m=16$}\\ 
\end{tabular}
\caption{Comparisons of multiplexing decoder performance for a [\![512,8]\!] equal-block ($16\times 16$) HGP code obtained from the symmetric construction with $r=n=16$ using various assignment strategies for $m=4$ and $m=16$. In both cases, the random, sudoku, and diagonal strategies are seen to be effectively equivalent to the no-multiplexing case, even at low erasure rates.}
\label{fig:n512_sym_m4_m16}
\end{figure*}

To compare the effectiveness of these strategies, we have simulated their performance for several codes at different multiplexing values as shown in Fig.~\ref{fig:n320_m8} and Fig.~\ref{fig:n512_sym_m4_m16}. To understand these results, the case with no-multiplexing ($m=1$) is used as the baseline. An assignment strategy is considered good if its failure rate is not significantly worse than the $m=1$ case. Interestingly, our numerical simulations consistently show that the performances of some strategies (random, sudoku, and diagonal) are almost equivalent to or even exceed the $m=1$ case, even at high multiplexing values. However, the row-col and stabilizer strategies are never seen to be effective in our results.

Fig.~\ref{fig:n320_m8} shows an example of a code where the diagonal strategy consistently outperforms all other strategies, even the no-multiplexing case, and even at low erasure rates. This result is significant because even though multiplexing reduces the number of required physical resources, it is possible to improve the decoding performance while doing so. In fact, an analysis of these results reveals that the diagonal strategy yields fewer logical errors than the no-multiplexing case at the same physical erasure rate. This appears to be a feature of the structure of the logical operators in the randomly generated code used in this simulation, even though the strategy was not designed with this in mind. This also explains the gap between the sudoku and diagonal strategies, both of which have similar amounts of decoder failures but differ with respect to logical errors. These results show that strategies designed to avoid decoder failures can have comparable (or even favorable) performance relative to the no-multiplexing case.

Although not identical, we observe similar performance for a larger HGP code, as shown in Fig.~\ref{fig:n512_sym_m4_m16}. The random, sudoku, and diagonal strategies have nearly identical performance to the no-multiplexing case regardless of the chosen multiplexing number. (Simulations include $m\in\{2,4,8,16\}$, although only plots for $m=4$ and $m=16$ are shown.) Furthermore, these results hold consistently at a low erasure rate, which is the regime of practical interest. This is significant because it implies there is no loss in performance when multiplexing, even though fewer physical resources are required, provided the assignment strategy is adapted to the decoder. If an ML decoder were used (e.g., Gaussian elimination rather than peeling + pruned peeling + VH), a gap is expected between the multiplexing and no-multiplexing cases. However, given that the combined decoder is a faster, more efficient alternative to a true ML decoder for HGP codes, these results are very promising.

All the programs we used to simulate multiplexed quantum communication with HGP codes are available here~\cite{Connolly_Python_implementation_of_2024}.

\section{Discussion and Conclusion}
\label{sec_conc}
We proposed three error-corrected quantum information processing scenarios for quantum memory storage and communication with quantum multiplexing over an erasure channel. We have shown that quantum multiplexing can improve throughput or resilience to errors, easing the bottleneck in quantum systems. This work can be adapted to error-corrected quantum communication ~\cite{fowler2010surface} with quantum interconnects, quantum repeaters, and multimode quantum memory~\cite{afzelius2009multimode}.

For multiplexed quantum communication, if multiple qubits in a single code word are encoded into the same photon, a correlation of errors in those qubits will be introduced. The simulation results show that it leads to an increase in the logical error rate. We showed that this performance gap can be significantly mitigated by introducing a code-aware (or decoder-aware) strategy to assign qubits to photons, which exploits code structure. For surface codes in particular, both randomness and distance maximization are important factors for achieving this. For HGP codes with the VH decoder, minimizing decoder failures was found to be the most important factor.

These techniques can also be exploited to benefit other families of codes and decoders.
Furthermore, it is possible to deal with the gap between the multiplexing and no-multiplexing cases by increasing the code size. We have also shown that it is possible to introduce biased error by using a stabilizer-based assignment strategy. In the special case of the diagonal strategy for the HGP code of Fig.~\ref{fig:n320_m8}, we see that the photon-correlated errors offer an improvement over the no-multiplexing case. In this example, the improvement can be explained by the fact that the diagonal strategy reduces logical errors in addition to decoder failures. Furthermore, this shows the existence of strategies that improve over no-multiplexing despite the fact that fewer resources are used.

Even though a linear-time ML decoder has not yet been discovered for generic HGP codes as it has been for surface codes, the use of HGP codes with quantum multiplexing should not be overlooked. Unlike surface codes, which have fixed dimension $2$, HGP codes can be chosen so that code dimension $k$ increases linearly with code length $n$. This can be significant for applications using increasingly long codes since the code rate need not approach $0$ in the HGP case. Furthermore, while an ML decoder is ideal, non-ML decoders are often good enough for error correction in the regime of practical interest. The speed-up gained by using a more efficient decoder can offset new errors that might arise when using a slower decoder. Our numerical results for HGP codes show that decoder-aware strategies enable us to gain all of the benefits of quantum multiplexing without sacrificing any additional performance. This, combined with the throughput advantage HGP codes have over the surface code, is a very promising practical result.

Although we propose several promising candidates, the optimal assignment strategy for both surface codes and HGP codes is still unknown. Furthermore, in actual communication with quantum multiplexing, various errors may occur when converting qubits in the quantum processor to photons, measuring stabilizers, and substituting erased qubits with mixed states. How to deal with these errors is a practically important next question.

Multiplexing could also be used for qubit $\rightarrow$ qudit encodings in non-photonic systems where loss errors are not the dominant source of noise. In these cases error locations may not be known and so knowledge of the assignment strategy could be used to inform decoding. 

\section*{Acknowledgements}
SN acknowledges Dan Browne, Antonio deMarti iOlius, and Hon Wai Lau for valuable discussions throughout this project. This work was supported by JSPS KAKENHI Grant Number JP22KJ1436, JP21H04880, the MEXT Quantum Leap Flagship Program (MEXT Q-LEAP) Grant Number JPMXS0118069605, the JST Moonshot R\&D Grant Number JPMJMS2061 and JPMJMS226C, and a travel budget from the National Institute of Informatics.

\bibliographystyle{quantum}
\bibliography{main_bib}

\appendix
\section{Encoding \texorpdfstring{$2^2$} \, dimensional quantum information in a photon}
\label{app:qmencoder2}

Here we give the transition of states in the circuit shown in Fig.~\ref{fig:qm_encoder}~(B) and (C).
\subsection{State transition for Fig.~\ref{fig:qm_encoder}~(B)}

\begin{widetext}
\begin{eqnarray*}
    && (\alpha\ket{HS} + \beta\ket{HL} + \gamma \ket{VS} + \delta\ket{VL}) \otimes \ket{+} \otimes \ket{+}\\
    &\xrightarrow{CX_{\textrm{m}_{1},\textrm{p}}}& \left( \frac{1}{\sqrt{2}}
    (\alpha \ket{HS} + \beta \ket{HL} + \gamma \ket{VS} + \delta\ket{VL}) \ket{0} + (\alpha \ket{VS} + \beta \ket{VL} + \gamma \ket{HS} + \delta\ket{HL})\ket{1}) \right)\\
    && \otimes \ket{+} \cdots\ctext{1} \\
    & \xrightarrow{\textrm{SW}}&
    \left( \frac{1}{\sqrt{2}}
    (\alpha \ket{HS} + \beta \ket{VS} + \gamma \ket{HL} + \delta\ket{VL}) \ket{0} + (\alpha \ket{HL} + \beta \ket{VL} + \gamma \ket{HS} + \delta\ket{VS})\ket{1}) \right)\\
    && \otimes \ket{+} \cdots\ctext{2} \\
    & \xrightarrow{CX_{\textrm{m}_{2}, \textrm{p}}}&
    \frac{1}{\sqrt{2}}( (\alpha \ket{HS} + \beta \ket{VS} + \gamma \ket{HL} + \delta\ket{VL})\ket{00} + (\alpha \ket{VS} + \beta \ket{HS} + \gamma \ket{VL} + \delta\ket{HL})\ket{01} \\
    &&+ (\alpha \ket{HL} + \beta \ket{VL} + \gamma \ket{HS} + \delta\ket{VS}) \ket{10}  + (\alpha \ket{VL} + \beta \ket{HL} + \gamma \ket{VS} + \delta\ket{HS})\ket{1})\ket{11})\\
    &&\cdots\ctext{3}
\end{eqnarray*}
\end{widetext}

To obtain the state $\alpha\ket{00} + \beta\ket{01} + \gamma \ket{10} + \delta \ket{11}$ one can apply the $II/IX/XI/XX$ gates on the matter qubits upon a corresponding photon measurement outcome $HS/VS/HL/VL$, respectively.

\subsection{State transition for Fig.~\ref{fig:qm_encoder}~(C)}
We use the $\alpha\ket{00} + \beta\ket{01} + \gamma \ket{10} + \delta \ket{11}$ as the initial state of the matter qubit $m_1,m_2$. The state transition is:
\begin{eqnarray*}
    &&\ket{H} \otimes (\alpha\ket{00} + \beta\ket{01} + \gamma \ket{10} + \delta \ket{11})\\
    & \xrightarrow{CX_{\textrm{m}_{2},\textrm{p}}} & \alpha\ket{H00} + \beta\ket{V01} \\
    && + \gamma \ket{H10} + \delta\ket{V11}\cdots\ctext{1}\\
    & \xrightarrow{\textrm{SW}}& \alpha\ket{HS00} + \beta\ket{HL01} \\
    && +\gamma\ket{HS10}+ \delta\ket{HL11} \cdots\ctext{2}\\
    & \xrightarrow{CX_{\textrm{m}_{1}, \textrm{p}}} & \alpha\ket{HS00} + \beta\ket{HL01} \\
    && +\gamma\ket{VS10}+ \delta\ket{VL11} \cdots\ctext{3}
\end{eqnarray*}
Then one can measure matter qubits in $X$-basis to obtain the desired state.

\section{Encoding \texorpdfstring{$2^k$} \, dimensional quantum information in a photon}
\label{app:qmencoder}
\begin{figure*}
    \centering
    \includegraphics[width=1.9\columnwidth]{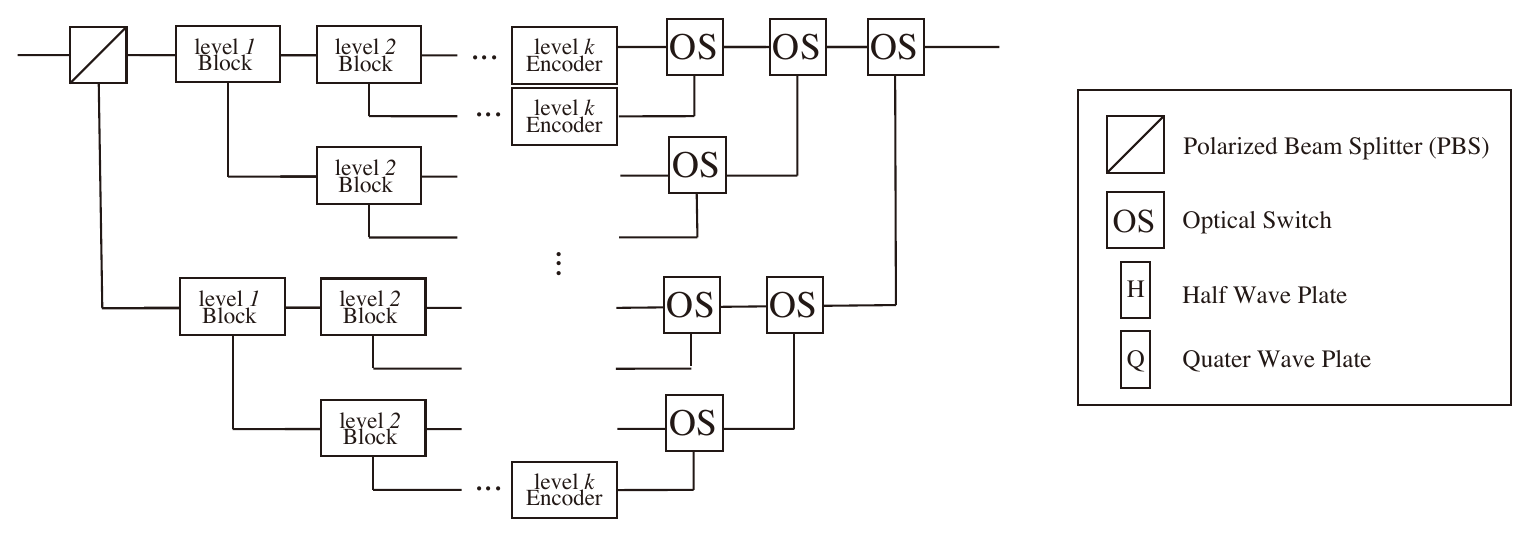}
    \caption{Optical circuit for preparing a photon with $2^k$ dimensional time-bin information.}
    \label{fig:level_k}
\end{figure*}

\begin{figure}
    \centering
    \includegraphics[width=1\columnwidth]{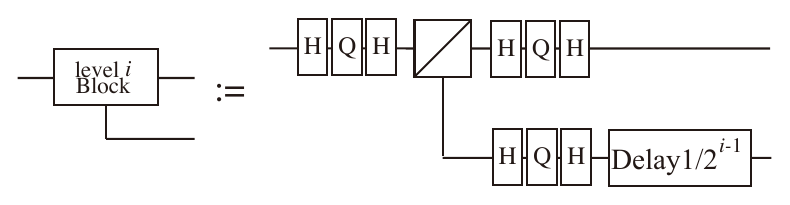}
    \caption{Level $i$ block used in Fig.~\ref{fig:level_k}. It splits a single mode into two modes with a new component of time-bin DOF introduced to the state.}
    \label{fig:level_i}
\end{figure}

Here we show the circuit for encoding $2^k$ dimensional quantum information into time-bin in one photon in Fig.~\ref{fig:level_k}, where the level $i$ block is defined as in Fig.~\ref{fig:level_i}. In the first half of the circuit, a level $i$ block is used to introduce a new component of the time-bin by applying a delay. Then, in the middle of the circuit, each state corresponds to a different mode. In the second half of the circuit, multiplexed photons are output to one mode by applying the optical switches, forming a complete binary tree. This encoding circuit can prepare a photon with $2^k$ dimensional time-bin and $2$ dimensional polarization in linear time with respect to $k$.

\section{Assignment strategies for surface codes}
\label{app:strategies_for_sc}
In this appendix, we describe the details and discussions on each assignment strategy for the surface code.

\textbf{strategy i and ii: pair with minimum and maximum distance} 
The Manhattan distance between qubits within the same photon is crucial when considering the impact of correlated errors. These strategies are deterministic and can be realized with simple calculations. 

\textbf{strategy iii: random} 
Errors with strong correlation are similar to burst errors in classical communication in the sense that the errors have spatial locality. This locality of errors can be addressed by classical error-correcting codes using two methods. The first method treats multiple bits as a single symbol (an element of a finite field), such as BCH codes and Reed-Solomon codes~\cite{wicker1999reed}. Thanks to the high correctability of burst errors, Reed-Solomon codes are used in many classical systems, including QR codes~\cite{qrc1995}, CDs, and satellite communications. Another method is the interleaving~\cite{proakis1987digital} technique. Interleaving eliminates locality by permuting the rows and columns of the code's generator matrix. There is also a method to apply interleaving to QECCs\cite{kawabata2000quantum}. Inspired by interleaving, we have constructed two strategies for quantum multiplexing with randomness.

\textbf{strategy iv: random + threshold} 
The fourth strategy is a modified version of the third strategy. The pseudo-code is shown below in Algorithm~\ref{alg:randomplusthreshold}. The flow of the algorithm is as follows: A ``threshold'' $T$ is set as $2/d - 1$, which is the maximum distance between two qubits in the $[\![2d^2, 2, d]\!]$ surface code. This value will be used to check that the distance between the qubits in the same photon is high enough. Then, it randomly assigns qubits for each photon while respecting the distance threshold. It randomly selects the first qubit of the photon, then it randomly selects a qubit again and takes it as a candidate to assign it to this photon. When the distance between the candidate qubit and the qubit(s) already in the photon is greater than the threshold, the qubit is accepted, and when it is less, it is rejected. This procedure is repeated until the photon has been assigned the appropriate number of qubits. If no qubit satisfies the threshold, then the threshold value is lowered by one. This ensures that the algorithm will always terminate. By repeating this process, we can assign all the qubits to photons. This strategy is designed to leverage randomness while also increasing the distance between qubits in the same photon. 

This algorithm requires calculating the distance between two qubits, which is easy for the surface code thanks to the fact that the distance between qubits in the grid representation is defined by the taxicab metric (the Manhattan distance). Note that this and other assignment strategies can still be applied even if the number $m$ of qubits per photon is not a divisor of the total number of qubits. In this case, we allow for a final ``remainder'' photon containing fewer than $m$ qubits.
\begin{algorithm}[ht]
    \caption{Strategy iv. random + threshold}
    \label{alg:randomplusthreshold}
    \KwData{$P=\{p_i\}$ (the set of photons) where initially $p_i=\{\emptyset\}$ (the set of qubits to be encoded in the $i^{\text{th}}$ photon), $Q=\{q_j\}$ (the list of all physical qubits in the code), and the multiplexing number $m$ of qubits in each photon.}
    \KwResult{$P=\{p_i\}$ (a set of sets of qubits assigned to the $i^{\text{th}}$ photon).}

    Initialize the threshold with $T$ := $\frac{d}{2} - 1$\;
    \For{photon $p_i\in P$}{
        Pick a qubit $q_j\in Q$ randomly\;
        Move $q_j$ from $Q$ to $p_i$\;
        \While{$|p_i|<m$}{
            \While{$|p_i|<m$ and $Q\neq\emptyset$}{
                Pick a candidate qubit $q_k\in Q$ at random\;
                \eIf{the distance between $q_k$ and all qubits in $p_i$ is greater than $T$}{
                    Move $q_k$ from $Q$ to $p_i$\;
                }{
                    Move $q_k$ from $Q$ to a waiting list $Q'$\;
                }
            }
            Move all qubits in $Q'$ to $Q$\;
            Update $T$ := $T - 1$\;        
        }
    }
    Return $P$;
\end{algorithm}

\textbf{Strategy v: Stabilzier} Error correction on the surface code is always considered up to multiplication by a stabilizer. This suggests that it may be useful to define photons using the qubit-support of a stabilizer check. Since the stabilizer generators for the surface code correspond to squares and crosses in the lattice, they have weight 4. On a $d\times d$ lattice, if $d$ is divisible by 4, it will always be possible to partition the lattice into squares and crosses. In this perspective, the L-shapes used in the minimum-distance strategy can be thought of as ``half-stabilizers'' in the lattice. Since the usual strategy for converting an erasure problem into an error correction problem involves assigning erased qubits Pauli errors randomly, this stabilizer assignment strategy uses a mix of $Z$ and $X$-type stabilizer generators from both squares and crosses. In this case, qubits are equally partitioned into the two types of stabilizers by tiling the lattice with alternating diagonal lines of squares and crosses. 

\section{Peeling Decoder}
\label{app:peeling}
The \textit{peeling decoder} refers to a linear-complexity erasure decoding algorithm originally designed for classical codes~\cite{luby2001efficient}. This algorithm corrects an erasure error by examining the subgraph of the Tanner graph corresponding to erased bits, whereby degree-1 check nodes in this subgraph give perfect information about adjacent bit nodes. Although not a maximum-likelihood decoder, the peeling decoder works well for codes with sparse Tanner graphs, such as LDPC codes. Because this algorithm only uses the Tanner graph, it can be directly applied to CSS codes as well. In this section, we briefly summarize the peeling decoder algorithm and several of its variations, beginning with a review of the classical erasure setting.

\subsection{Peeling Algorithm for Classical Codes}
\label{app:peeling_classical}
For a classical code, an erasure error on a codeword can be modeled as the loss of a known subset of bits. By assigning these erased bits the values 0 or 1 at random and then making a syndrome measurement, the erasure correction problem can be converted into an error correction problem. Unlike standard error correction, we make the additional assumption that non-erased bits do not have errors. Hence, it is sufficient to consider error correction using the subcode corresponding only to erased bits. In terms of the Tanner graph, this is equivalent to considering the subgraph induced by the erasure (consisting of the subset of erased bit-nodes and any adjacent check-nodes). The peeling algorithm is defined in terms of this erasure-induced subgraph.

\begin{figure}[ht]
    \centering
    {
    \begin{tikzpicture}[baseline]
    
    \node (1) [circle,draw,inner sep=3pt,line width=1,opacity=0.5,color=gray] at (0,3) {};
    \node (2) [circle,draw,inner sep=3pt,line width=1,opacity=0.5,color=gray] at (0,2.5) {};
    \node (3) [circle,draw,inner sep=3pt,line width=1] at (0,2) {};
    \node (4) [circle,draw,inner sep=3pt,line width=1] at (0,1.5) {};
    \node (5) [circle,draw,inner sep=3pt,line width=1] at (0,1) {};
    \node (6) [circle,draw,inner sep=3pt,line width=1,opacity=0.5,color=gray] at (0,0.5) {};
    \node (7) [circle,draw,inner sep=3pt,line width=1,opacity=0.5,color=gray] at (0,0) {};
    
    \node (a) [rectangle,draw,inner sep=4.5pt,line width=1] at (1.25,2.25) {};
    \node (b) [rectangle,draw,inner sep=4.5pt,line width=1,fill=red] at (1.25,1.5) {};
    \node (c) [rectangle,draw,inner sep=4.5pt,line width=1] at (1.25,0.75) {};

    \path (1) edge[line width=1.5,opacity=0.5,color=gray] (c);
    
    \path (2) edge[line width=1.5,opacity=0.5,color=gray] (b);
    
    \path (3) edge[line width=1.5] (b);
    \path (3) edge[line width=1.5] (c);
    
    \path (4) edge[line width=1.5] (a);
    
    \path (5) edge[line width=1.5] (a);
    \path (5) edge[line width=1.5] (c);
    
    \path (6) edge[line width=1.5,opacity=0.5,color=gray] (a);
    \path (6) edge[line width=1.5,opacity=0.5,color=gray] (b);
    
    \path (7) edge[line width=1.5,opacity=0.5,color=gray] (a);
    \path (7) edge[line width=1.5,opacity=0.5,color=gray] (b);
    \path (7) edge[line width=1.5,opacity=0.5,color=gray] (c);
    \end{tikzpicture}
    }
    {
    \begin{tikzpicture}[baseline]
    
    \node (1) [circle,draw,inner sep=3pt,line width=1,opacity=0.5,color=gray] at (0,3) {};
    \node (2) [circle,draw,inner sep=3pt,line width=1,opacity=0.5,color=gray] at (0,2.5) {};
    \node (3) [circle,draw,inner sep=3pt,line width=1,opacity=0.5,color=gray] at (0,2) {};
    \node (4) [circle,draw,inner sep=3pt,line width=1] at (0,1.5) {};
    \node (5) [circle,draw,inner sep=3pt,line width=1] at (0,1) {};
    \node (6) [circle,draw,inner sep=3pt,line width=1,opacity=0.5,color=gray] at (0,0.5) {};
    \node (7) [circle,draw,inner sep=3pt,line width=1,opacity=0.5,color=gray] at (0,0) {};
    
    \node (a) [rectangle,draw,inner sep=4.5pt,line width=1] at (1.25,2.25) {};
    \node (b) [rectangle,draw,inner sep=4.5pt,line width=1,opacity=0.5,color=gray] at (1.25,1.5) {};
    \node (c) [rectangle,draw,inner sep=4.5pt,line width=1,fill=red] at (1.25,0.75) {};

    \path (1) edge[line width=1.5,opacity=0.5,color=gray] (c);
    
    \path (2) edge[line width=1.5,opacity=0.5,color=gray] (b);
    
    \path (3) edge[line width=1.5,opacity=0.5,color=gray] (b);
    \path (3) edge[line width=1.5,opacity=0.5,color=gray] (c);
    
    \path (4) edge[line width=1.5] (a);
    
    \path (5) edge[line width=1.5] (a);
    \path (5) edge[line width=1.5] (c);
    
    \path (6) edge[line width=1.5,opacity=0.5,color=gray] (a);
    \path (6) edge[line width=1.5,opacity=0.5,color=gray] (b);
    
    \path (7) edge[line width=1.5,opacity=0.5,color=gray] (a);
    \path (7) edge[line width=1.5,opacity=0.5,color=gray] (b);
    \path (7) edge[line width=1.5,opacity=0.5,color=gray] (c);
    \end{tikzpicture}
    }
    {
    \begin{tikzpicture}[baseline]
    
    \node (1) [circle,draw,inner sep=3pt,line width=1,opacity=0.5,color=gray] at (0,3) {};
    \node (2) [circle,draw,inner sep=3pt,line width=1,opacity=0.5,color=gray] at (0,2.5) {};
    \node (3) [circle,draw,inner sep=3pt,line width=1,opacity=0.5,color=gray] at (0,2) {};
    \node (4) [circle,draw,inner sep=3pt,line width=1] at (0,1.5) {};
    \node (5) [circle,draw,inner sep=3pt,line width=1,opacity=0.5,color=gray] at (0,1) {};
    \node (6) [circle,draw,inner sep=3pt,line width=1,opacity=0.5,color=gray] at (0,0.5) {};
    \node (7) [circle,draw,inner sep=3pt,line width=1,opacity=0.5,color=gray] at (0,0) {};
    
    \node (a) [rectangle,draw,inner sep=4.5pt,line width=1,fill=red] at (1.25,2.25) {};
    \node (b) [rectangle,draw,inner sep=4.5pt,line width=1,opacity=0.5,color=gray] at (1.25,1.5) {};
    \node (c) [rectangle,draw,inner sep=4.5pt,line width=1,opacity=0.5,color=gray] at (1.25,0.75) {};

    \path (1) edge[line width=1.5,opacity=0.5,color=gray] (c);
    
    \path (2) edge[line width=1.5,opacity=0.5,color=gray] (b);
    
    \path (3) edge[line width=1.5,opacity=0.5,color=gray] (b);
    \path (3) edge[line width=1.5,opacity=0.5,color=gray] (c);
    
    \path (4) edge[line width=1.5] (a);
    
    \path (5) edge[line width=1.5,opacity=0.5,color=gray] (a);
    \path (5) edge[line width=1.5,opacity=0.5,color=gray] (c);
    
    \path (6) edge[line width=1.5,opacity=0.5,color=gray] (a);
    \path (6) edge[line width=1.5,opacity=0.5,color=gray] (b);
    
    \path (7) edge[line width=1.5,opacity=0.5,color=gray] (a);
    \path (7) edge[line width=1.5,opacity=0.5,color=gray] (b);
    \path (7) edge[line width=1.5,opacity=0.5,color=gray] (c);
    \end{tikzpicture}
    }
    {
    \begin{tikzpicture}[baseline]
    
    \node (1) [circle,draw,inner sep=3pt,line width=1,opacity=0.5,color=gray] at (0,3) {};
    \node (2) [circle,draw,inner sep=3pt,line width=1,opacity=0.5,color=gray] at (0,2.5) {};
    \node (3) [circle,draw,inner sep=3pt,line width=1,opacity=0.5,color=gray] at (0,2) {};
    \node (4) [circle,draw,inner sep=3pt,line width=1,opacity=0.5,color=gray] at (0,1.5) {};
    \node (5) [circle,draw,inner sep=3pt,line width=1,opacity=0.5,color=gray] at (0,1) {};
    \node (6) [circle,draw,inner sep=3pt,line width=1,opacity=0.5,color=gray] at (0,0.5) {};
    \node (7) [circle,draw,inner sep=3pt,line width=1,opacity=0.5,color=gray] at (0,0) {};
    
    \node (a) [rectangle,draw,inner sep=4.5pt,line width=1,opacity=0.5,color=gray] at (1.25,2.25) {};
    \node (b) [rectangle,draw,inner sep=4.5pt,line width=1,opacity=0.5,color=gray] at (1.25,1.5) {};
    \node (c) [rectangle,draw,inner sep=4.5pt,line width=1,opacity=0.5,color=gray] at (1.25,0.75) {};

    \path (1) edge[line width=1.5,opacity=0.5,color=gray] (c);
    
    \path (2) edge[line width=1.5,opacity=0.5,color=gray] (b);
    
    \path (3) edge[line width=1.5,opacity=0.5,color=gray] (b);
    \path (3) edge[line width=1.5,opacity=0.5,color=gray] (c);
    
    \path (4) edge[line width=1.5,opacity=0.5,color=gray] (a);
    
    \path (5) edge[line width=1.5,opacity=0.5,color=gray] (a);
    \path (5) edge[line width=1.5,opacity=0.5,color=gray] (c);
    
    \path (6) edge[line width=1.5,opacity=0.5,color=gray] (a);
    \path (6) edge[line width=1.5,opacity=0.5,color=gray] (b);
    
    \path (7) edge[line width=1.5,opacity=0.5,color=gray] (a);
    \path (7) edge[line width=1.5,opacity=0.5,color=gray] (b);
    \path (7) edge[line width=1.5,opacity=0.5,color=gray] (c);
    \end{tikzpicture}
    }
    {
    \begin{tikzpicture}[baseline]
        \node (start) at (0,0) {};
        \node (end) at (6,0) {};
        \path (start) edge[line width=1.5,-stealth,"$t$"] (end);
        \node (H) at (3,-1) {\small$H=\begin{bmatrix}0&0&0&1&1&1&1\\0&1&1&0&0&1&1\\1&0&1&0&1&0&1\end{bmatrix}$};
    \end{tikzpicture}
    }
    \caption{Example of the peeling algorithm applied to an erasure-induced subgraph of the Tanner graph of the classical code $C=\text{Ker}(H)$, where circle-nodes denote bits and square-nodes denote checks. Gray edges and nodes are not included in the erasure. Red squares indicate a dangling check (degree 1 check-node in the subgraph). At each time step, a dangling check and adjacent erased bit are removed from the erasure, until the erasure is empty.}
    \label{fig:peeling_decoder_example}
\end{figure}
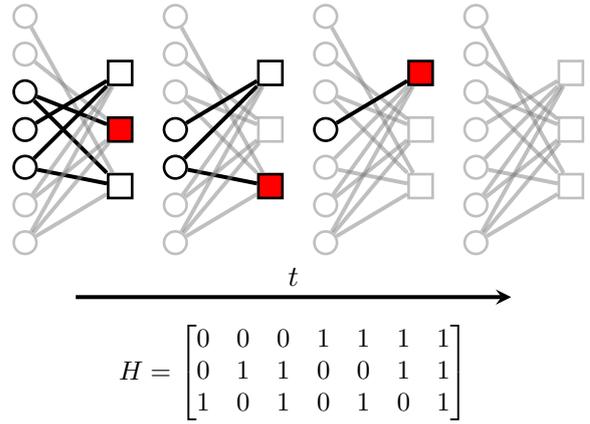

A check is said to be \textit{dangling} if it has degree 1 in the subgraph (i.e. a dangling check is adjacent to exactly one erased bit). Recall that each check-node corresponds to a position in the syndrome vector for the randomly selected erasure-supported error. The value of the syndrome bit corresponding to a dangling check gives perfect information about the error on the adjacent erased bit. Based on this, the error on this bit can be corrected and then removed from the original set of erased bits, thus shrinking the erasure-induced subgraph and possibly introducing new dangling checks. The peeling decoder functions by performing a sequence of partial corrections, one erased bit at a time, hence "peeling" the subgraph until no dangling checks remain. This process is shown visually in Fig.~\ref{fig:peeling_decoder_example} and briefly summarized in Algorithm~\ref{alg:peeling_decoder}.

\begin{algorithm}[ht]
    \caption{Peeling Algorithm}
    \label{alg:peeling_decoder}
    \KwData{A code $\text{Ker}(H)$ with Tanner graph $G$, a set of erased bits $E$, and a syndrome vector $s$.}
    \KwResult{A predicted error $\hat{e}\subseteq E$ such that $H\hat{e}=s$, or {\bf Failure}.}

    Initialize $\hat{e}=\emptyset$\;
    \While{$E\neq\emptyset$}{
        Compute erasure subgraph $G_E\subseteq G$\;
        \eIf{$\exists$ dangling check $s_i\in G_E$}{
            \eIf{$s_i$ is unsatisfied}{
                $\exists$ error on adjacent bit $b_j\in E$\;
                Flip bit $b_j$, update syndrome $s$\;
                Update $\hat{e} := \hat{e}\cup\{b_j\}$\;
            }{
                No error on adjacent bit $b_j$\;
            }
            Update $E := E\setminus\{b_j\}$\;
        }{
            Return {\bf Failure}\;
        }
    }
    Return $\hat{e}$\;
\end{algorithm}

The decoding is successful if all erased bits have been corrected. A decoding failure occurs when all dangling checks have been peeled, but the remaining erasure is nonempty (i.e. all remaining checks in the subgraph have degree 2 or higher). Such a configuration is referred to as a \textit{stopping set}. If an erasure pattern contains a stopping set, then the peeling algorithm will fail to find a correction. In particular, because the algorithm may not even return to the code space, this shows that peeling is not a maximum-likelihood decoder.

\subsection{Peeling Algorithm for CSS Codes}
\label{app:peeling_CSS}
Erasure correction for a quantum code is modeled similarly to the classical case, with an erasure error on a codeword corresponding to the loss of a known subset of qubits. As in the classical case, erasure correction can be converted into error correction, with the modified rule that erased qubits are assigned Pauli errors in $\{I,X,Z,Y\}$ at random in the quantum case. For a CSS code, errors can be corrected by applying the peeling algorithm two times, once using the classical Tanner graph for $H_Z$ and once again for $H_X$. Since $X$- and $Z$-type Pauli errors are corrected independently, the same initial erasure pattern is used both times.

\subsection{Peeling Algorithm for Surface Codes}
\label{app:peeling_surface}
The \textit{surface code peeling decoder} refers to a generalization of this algorithm adapted to surface codes~\cite{delfosse2020linear}, which uses additional information about stabilizers in the code. Before applying the standard peeling algorithm, the modified algorithm first computes a certain acyclic subgraph of the usual erasure-induced subgraph. By leveraging stabilizer equivalences, it is sufficient to apply the peeling algorithm only to this acyclic subgraph to correct the entire erasure error; the random values assigned to erased qubits not included in this subgraph are assumed to be correct. The advantage here is that an acyclic graph does not contain stopping sets; the peeling algorithm will always successfully terminate with a predicted erasure correction when applied to the acyclic subgraph in question. Hence, unlike the standard peeling decoder, the surface code peeling decoder is maximum-likelihood.

It remains to comment on how we compute this acyclic subgraph of the erasure-induced subgraph of the Tanner graph for the surface code. To explain this, we consider the usual depiction of a distance $d$ surface code on a $d\times d$ lattice, whereby qubits are identified with edges in the lattice and $X$- and $Z$-type stabilizer checks are identified with vertices and plaquettes, respectively. That is say, the $H_X$-computed syndrome for $Z$-type Pauli errors on qubits is visualized by the subset of vertices corresponding to unsatisfied $X$-checks. A similar visualization for $X$-type Pauli errors is possible using the dual graph of this lattice picture. In the context of an erasure error, a subset of erased qubits is visualized by a corresponding set of erased edges in the surface code lattice. This erasure can also be thought of as the subgraph of the lattice consisting of erased qubit-edges and any vertices adjacent to these edges (not to be confused with the related erasure-induced subgraph of the Tanner graph).

After assigning erased qubits Pauli errors at random, as usual, we consider the correction of $Z$- and $X$-type errors independently. In the $Z$-error case, the syndrome corresponding to the unsatisfied $X$-checks is a subset of the vertices in the erasure-induced subgraph of the lattice. The algorithm proceeds by computing a spanning tree of the erasure-induced subgraph of the lattice (or a spanning forest in the case of a disjoint subgraph). This spanning tree in the lattice also corresponds to a subgraph in the Tanner graph of $H_X$; each leaf in the spanning tree corresponds to a dangling check in the subgraph. In this way, we obtain the acyclic subgraph of the erasure-induced subgraph of the Tanner graph mentioned earlier. Any two spanning trees are equivalent up to multiplication by stabilizers, as are the predicted errors obtained via the peeling algorithm. The $X$-errors are corrected in exactly the same way, except using the dual lattice.

\begin{figure}[ht]
    \centering
    \includegraphics[width = 1 \columnwidth]{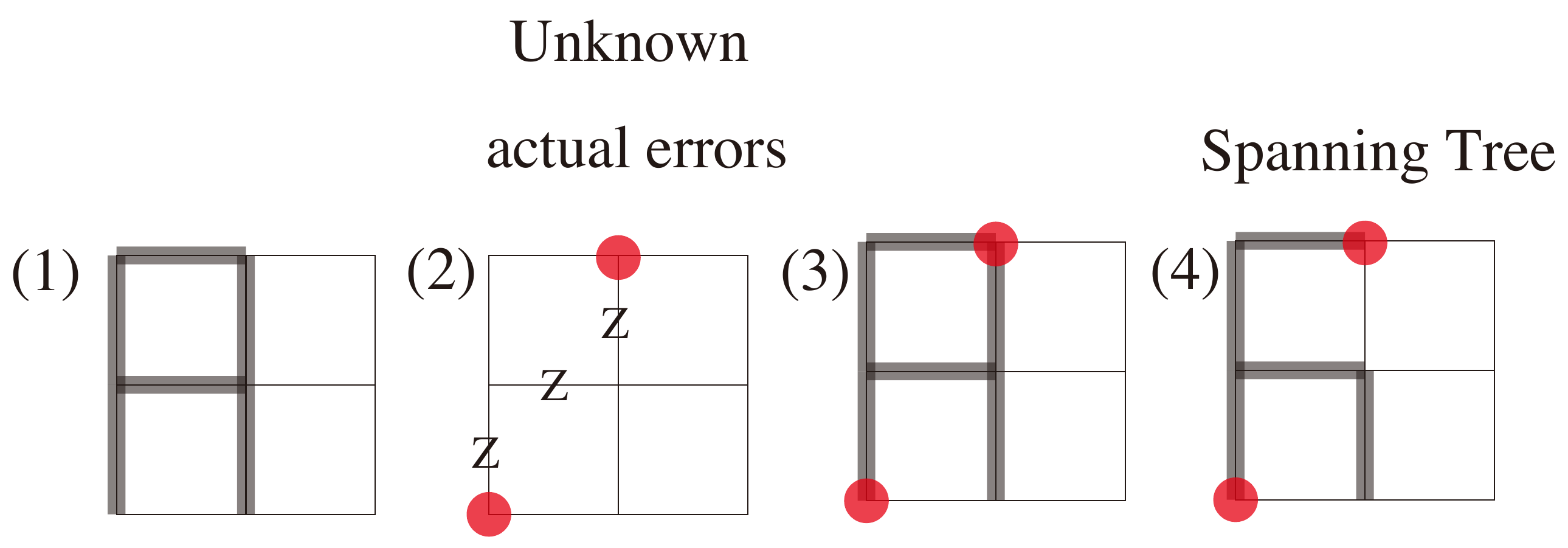}
    \caption{Illustration of the error correction process for an example of erasure errors with the surface code lattice. (1) Erased qubits are shown in bold grey lines. (2) Erasure errors are converted to random Pauli errors by replacing erased qubits with mixed states. The syndrome (indicated by the red vertices) is then computed by applying stabilizer measurements as explained in Sec.~\ref{sec_ER}. (3) Information seen by the decoding algorithm: erasure pattern and syndrome. (4) A spanning tree for the erasure pattern in the lattice is computed; this is identified with a corresponding acyclic subgraph of the Tanner graph. The \textit{surface code peeling decoder} then corrects the qubits one by one using this subgraph.}
    \label{fig:sc_peeling_step1}
\end{figure}

\begin{figure}[ht]
    \centering
    \includegraphics[width = 1 \columnwidth]{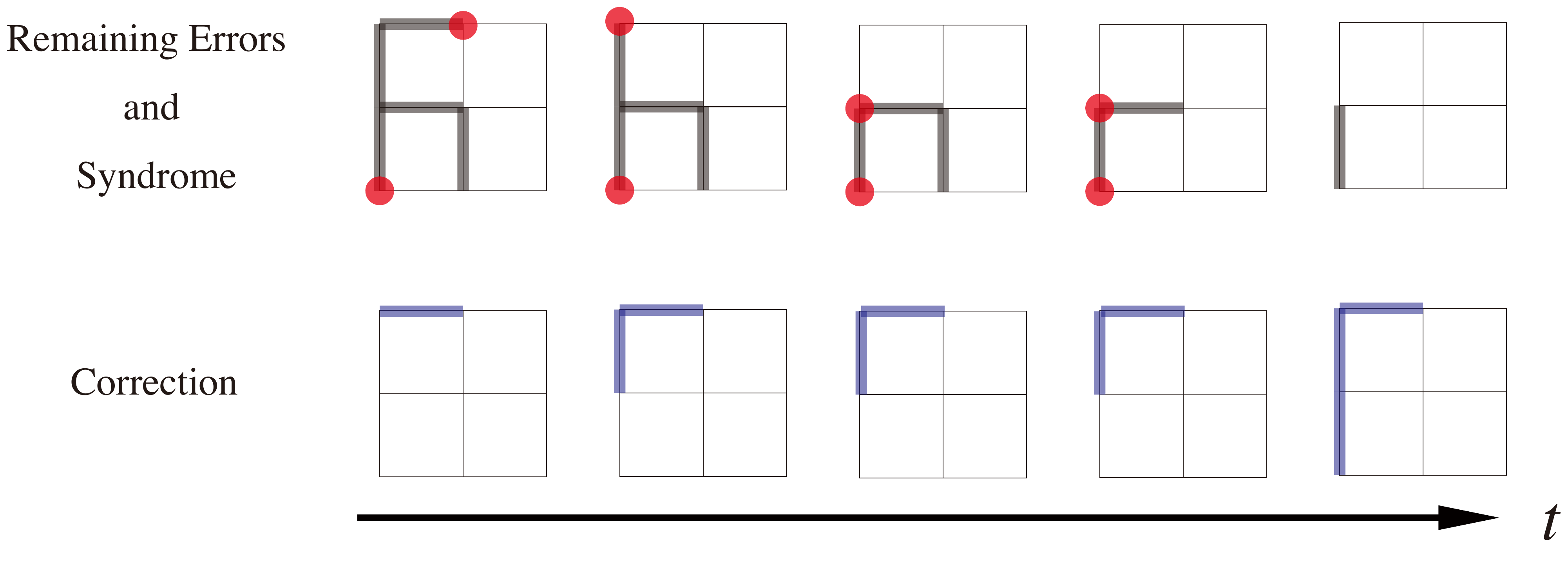}
    \caption{Peeling process to decode the erasure error pattern given in Fig.~\ref{fig:sc_peeling_step1}. Grey edges indicate the spanning tree in the surface code lattice and red vertices indicate the syndrome. The peeling algorithm is applied to the corresponding subgraph of the Tanner graph. Blue edges indicate corrections applied to qubits. Each time step denotes one iteration of the peeling algorithm, whereby the erasure is reduced by one qubit.}
    \label{fig:sc_peeling_step2}
\end{figure}

This modified peeling algorithm is a linear-complexity, maximum-likelihood decoder for the surface code. We use this algorithm in our numerical simulations for the surface code. Our implementation of the surface code peeling decoder is available in~\cite{Nishio_C_implementation_of_2024}. This process is briefly summarized in Fig.~\ref{fig:sc_peeling_step1} and \ref{fig:sc_peeling_step2}.

\subsection{Peeling Algorithm for HGP Codes}
\label{app:peeling_HGP}
The \textit{pruned peeling + VH decoder}~\cite{connolly2024fast} is yet another generalization of the peeling decoder adapted to the special case of HGP codes. As mentioned in Appendix~\ref{app:peeling_CSS}, because these are a type of CSS code, the standard peeling algorithm can be directly applied to HGP codes. However, this algorithm performs very poorly in practice, even for LDPC codes. This poor performance can be explained by the presence of stopping sets unique to HGP codes which have no analogue in the classical case. These stopping sets can be grouped into two types: \textit{stabilizer} and \textit{classical}, both of which cause the decoder to fail. The pruned peeling + VH decoder is designed to address these stopping sets.

\textit{Stabilizer stopping sets} occur when the erasure contains the qubit-support of an $X$- or $Z$-type stabilizer. Recall that the $X$- and $Z$-type stabilizers must commute, which is possible if and only if they overlap on an even number of qubits. Hence, in terms of the nodes denoting stabilizer generators in the Tanner graph, if we fix a node corresponding to an $X$-stabilizer generator and consider the set of qubits on which it is supported, each $Z$-stabilizer generator with overlapping qubit-support must overlap on an even number of qubits in order to satisfy the commutativity condition. Since the nodes denoting $Z$ stabilizer generators are exactly the check nodes in the Tanner graph of $H_Z$, these constitute a stopping set, as shown in the example of Fig.~\ref{fig:stabilizer_stopping_set}. In particular, there exist no dangling checks (which have degree 1) and hence this is a peeling decoder-stopping set. A similar relationship holds for $Z$-stabilizers in the Tanner graph of $H_X$.

    \begin{figure}[h!]
        \centering
        \begin{tikzpicture}[baseline]
        \node (nw) [inner sep=0] at (-3/2,3/2) {};
        \node (n) [inner sep=0] at (0,3/2) {};
        \node (ne) [inner sep=0] at (3/2,3/2) {};
        \node (w) [inner sep=0] at (-3/2,0) {};
        \node (c) [inner sep=0] at (0,0) {};
        \node (e) [inner sep=0] at (3/2,0) {};
        \node (sw) [inner sep=0] at (-3/2,-3/2) {};
        \node (s) [inner sep=0] at (0,-3/2) {};
        \node (se) [inner sep=0] at (3/2,-3/2) {};
        
        \path (nw) edge[line width=1,opacity=1] (ne);
        \path (w) edge[line width=1,dashed,opacity=1] (e);
        \path (sw) edge[line width=1,opacity=1] (se);
        \path (nw) edge[line width=1,opacity=1] (sw);
        \path (n) edge[line width=1,dashed,opacity=1] (s);
        \path (ne) edge[line width=1,opacity=1] (se);
        
        \node(n1) at (-3.6/2,1.5/2) {\rotatebox{0}{$n_1$}};
        \node(n2) at (-1.5/2,3.6/2) {\rotatebox{0}{$n_2$}};
        \node(r1) at (-3.6/2,-1.5/2) {\rotatebox{0}{$r_1$}};
        \node(r2) at (1.2/2,3.6/2) {\rotatebox{0}{$r_2$}};

        \node (-22) [circle,draw,inner sep=3pt,line width=1,opacity=1] at (-2/2,2/2) {};
        \node (-12) [circle,draw,inner sep=3pt,line width=1,opacity=1,fill=red] at (-1/2,2/2) {};
        \node (-21) [circle,draw,inner sep=3pt,line width=1,opacity=1] at (-2/2,1/2) {};
        \node (-11) [circle,draw,inner sep=3pt,line width=1,opacity=1,fill=red] at (-1/2,1/2) {};
        
        \node (12) [rectangle,draw,inner sep=4pt,line width=1,opacity=1,fill=cyan] at (1/2,2/2) {};
        \node (22) [rectangle,draw,inner sep=4pt,line width=1,opacity=1,fill=cyan] at (2/2,2/2) {};
        \node (11) [rectangle,draw,inner sep=4pt,line width=1,opacity=1,fill=cyan] at (1/2,1/2) {};
        \node (21) [rectangle,draw,inner sep=4pt,line width=1,opacity=1,fill=cyan] at (2/2,1/2) {};

        \node (-2-1) [rectangle,draw,inner sep=4pt,line width=1,opacity=1] at (-2/2,-1/2) {};
        \node (-1-1) [rectangle,draw,inner sep=4pt,line width=1,opacity=1,fill=orange] at (-1/2,-1/2) {};
        \node (-2-2) [rectangle,draw,inner sep=4pt,line width=1,opacity=1] at (-2/2,-2/2) {};
        \node (-1-2) [rectangle,draw,inner sep=4pt,line width=1,opacity=1] at (-1/2,-2/2) {};
        
        \node (1-1) [circle,draw,inner sep=3pt,line width=1,opacity=1,fill=red] at (1/2,-1/2) {};
        \node (2-1) [circle,draw,inner sep=3pt,line width=1,opacity=1,fill=red] at (2/2,-1/2) {};
        \node (1-2) [circle,draw,inner sep=3pt,line width=1,opacity=1] at (1/2,-2/2) {};
        \node (2-2) [circle,draw,inner sep=3pt,line width=1,opacity=1] at (2/2,-2/2) {};

        \node (-26) [circle,draw,inner sep=3pt,line width=1,opacity=1] at (-2/2,5/2) {};
        \node (-16) [circle,draw,inner sep=3pt,line width=1,opacity=1] at (-1/2,5/2) {};
        \node (16) [rectangle,draw,inner sep=4pt,line width=1,opacity=1] at (1/2,5/2) {};
        \node (26) [rectangle,draw,inner sep=4pt,line width=1,opacity=1] at (2/2,5/2) {};
        \path (-26) edge[line width=1,bend left=40,looseness=0.75] (16);
        \path (-26) edge[line width=1,bend left=40,looseness=0.75] (26);
        \path (-16) edge[line width=1,bend left=40,looseness=0.75] (16);
        \path (-16) edge[line width=1,bend left=40,looseness=0.75] (26);
        
        \node (-62) [circle,draw,inner sep=3pt,line width=1,opacity=1] at (-5/2,2/2) {};
        \node (-61) [circle,draw,inner sep=3pt,line width=1,opacity=1] at (-5/2,1/2) {};
        \node (-6-1) [rectangle,draw,inner sep=4pt,line width=1,opacity=1] at (-5/2,-1/2) {};
        \node (-6-2) [rectangle,draw,inner sep=4pt,line width=1,opacity=1] at (-5/2,-2/2) {};
        \path (-62) edge[line width=1,bend left=-40,looseness=0.75] (-6-1);
        \path (-62) edge[line width=1,bend left=-40,looseness=0.75] (-6-2);
        \path (-61) edge[line width=1,bend left=-40,looseness=0.75] (-6-1);
        \path (-61) edge[line width=1,bend left=-40,looseness=0.75] (-6-2);
        
        \path (-1-1) edge[line width=1,bend left=-40,looseness=0.75,color=orange, bend right] (-12);
        \path (-1-1) edge[line width=1,bend left=-40,looseness=0.75,color=orange, bend right] (-11);
        \path (1-1) edge[line width=1,bend left=-40,looseness=0.75,color=orange, bend left] (-1-1);
        \path (2-1) edge[line width=1,bend left=-40,looseness=0.75,color=orange, bend left] (-1-1);
        \path (11) edge[line width=1,bend left=40,looseness=0.75,color=cyan, bend right] (-11);
        \path (11) edge[line width=1,bend left=40,looseness=0.75,color=cyan, bend right] (1-1);
        \path (12) edge[line width=1,bend left=40,looseness=0.75,color=cyan, bend right] (-12);
        \path (12) edge[line width=1,bend left=40,looseness=0.75,color=cyan, bend right] (1-1);
        \path (21) edge[line width=1,bend left=40,looseness=0.75,color=cyan, bend right] (-11);
        \path (21) edge[line width=1,bend left=40,looseness=0.75,color=cyan, bend right] (2-1);
        \path (22) edge[line width=1,bend left=40,looseness=0.75,color=cyan, bend right] (-12);
        \path (22) edge[line width=1,bend left=40,looseness=0.75,color=cyan, bend right] (2-1);

        \end{tikzpicture}
        \caption{An example of a stabilizer stopping set for $H_z$ of the $2 \times 2$ surface code. The red circles indicate an erasure error, which consists of all of the qubits adjacent to the $X$-stabilizer denoted by the orange square. The cyan squares denote the $Z$-stabilizers adjacent to this set of erased qubits. Notice that each of these $Z$-stabilizers is adjacent to an even number of erased qubits, which is a consequence of the commutation relation between $X$- and $Z$-type stabilizers and also guarantees that there are no peelable degree-$1$ checks in the erasure subgraph.}
        \label{fig:stabilizer_stopping_set}
    \end{figure}

Such a stopping set can be modified by fixing a value at random for one qubit of the stabilizer and removing this qubit from the erasure. This reduces the degree of a single check-node in the erasure subgraph by 1, possibly introducing a dangling check and allowing the standard peeling algorithm to become unstuck. Removing a qubit from the erasure is equivalent to declaring the random mixed state on this qubit to be correct. This technique is valid for CSS codes because there exists a solution on the remaining erased qubits in the stabilizer-support such that the combined contribution to the error is at most a stabilizer. This procedure, known as \textit{pruned peeling}, is applicable to any CSS code, not just HGP codes.

\textit{Classical stopping sets} are another common type of peeling decoder stopping set which are only defined for HGP codes. These refer to patterns of erased qubits supported entirely on a single row or column in the HGP Tanner graph block structure of Fig.~\ref{fig:HGP_TannerGraph}. In the simplest case, a peeling decoder stopping set for one of the classical codes used in the HGP construction lifts to a classical stopping set for the HGP code. Furthermore, any HGP peeling decoder stopping set can be decomposed into a union of vertical and horizontal sets on the columns and rows of the Tanner graph; although we refer to these components as \textit{classical stopping sets}, a single component in isolation need not be a stopping set for the corresponding classical code.

The \textit{VH decoder} algorithm functions by ordering and efficiently solving each of these classical stopping sets in sequence, when possible, using the Gaussian decoder. The basic premise relies on the fact that, for a HGP code of length $N$, the component classical codes have length on the order of $\sqrt{N}$. Hence, even though Gaussian elimination (which has cubic complexity in the code length) is usually too slow for practical use, the complexity is reduced when restricted to a single classical stopping set. However, classical stopping sets often overlap (i.e. share a check-node in the Tanner graph), in which case the two stopping sets cannot be resolved independently without introducing some additional restrictions. By checking these conditions, the VH decoder attempts to find solutions for classical stopping sets which are compatible in these overlapping cases. If such a solution is found for each classical stopping set, these combine to give a solution for the HGP code. However, there exist erasure configurations where the VH decoder becomes stuck as well. In general, these will occur when there exist cycles of classical stopping sets in the erasure-induced subgraph. 

The \textit{pruned peeling + VH decoder} refers to the combination of these three strategies (standard peeling, correction of stabilizer stopping sets, and correction of classical stopping sets). For simplicity, we also use the term \textit{combined decoder} to refer to peeling + pruned peeling + VH. A stopping set for the combined decoder meets three conditions: there exist no remaining dangling checks; the remaining erasure does not cover the qubit-support of a stabilizer; remaining classical stopping sets form a cycle. Although these happen infrequently at a low erasure rate, an erasure pattern of this form will result in a decoder failure. These are distinct from logical errors, which can only be identified in numerical simulations where the decoding algorithm successfully terminates. The maximum-likelihood decoder always terminates, and thus, logical errors are the only source of failures. Although we make a distinction between these two possibilities, we will use the term \textit{error recovery failure} to refer to either a decoder failure or a non-decoder-failure logical error. A more detailed discussion of these differences is included in Appendix~\ref{app:HGP_decoder}.

The computational complexity of the combined decoder is dominated by the step applying cubic-complexity Gaussian elimination to classical stopping sets (peeling and pruned peeling both have linear complexity). The number of classical stopping sets is on the order of the number of rows and columns in the Tanner graph ($\sqrt{N}$ for an HGP code of length $N$). Furthermore, since classical stopping sets have a size of approximately $\sqrt{N}$, the effect of Gaussian elimination on a single classical stopping set contributes $O(N^{1.5})$ to the complexity. This becomes $O(N^2)$ across all classical stopping sets, establishing this algorithm as a quadratic complexity decoder for HGP codes.

\section{Additional Details for HGP Codes}
\label{app:HGP}
\subsection{Symmetric Constructions}
\label{app:HGP_construction}
Recall equations~\ref{eq:HGP_Hx} and \ref{eq:HGP_Hz} used to define the parity check matrices for HGP codes. The the sizes of the matrices $H_X$ and $H_Z$ obtained in this way are determined by the sizes of the input classical matrices $H_1$ and $H_2$. Hence, the number of qubits and stabilizer checks are controlled by the size of the input classical matrices; equations~\ref{eq:HGP_Hx_size} and \ref{eq:HGP_Hz_size} give the exact dimensions for $H_X$ and $H_Z$ obtained from matrices $H_1=[r_1\times n_1]$ and $H_2=[r_2\times n_2]$. Referring to the Tanner graph of Fig.~\ref{fig:HGP_TannerGraph}, the $n_1\times n_2$ and $r_1\times r_2$ blocks denote the qubit nodes and the $r_1\times n_2$ and $n_1\times r_2$ blocks denote the $X$- and $Z$-type stabilizer generators, respectively.

Choosing classical matrices of the same size ensures an equal number of stabilizer checks in the HGP code, but a biased code can also be constructed by using matrices of different sizes. Furthermore, using $H_2=H_1^T$ yields a symmetric construction for $H_X$ and $H_Z$ and guarantees that the two blocks of qubits in this product graph picture are squares of equal size. In our numerical simulations, we consider two types of HGP code construction: an \textit{equal block} case coming from the symmetric construction with $H_2=H_1^T$, whereby $r_2=n_1$ and $n_2=r_1$; and a \textit{non-equal block} case using different matrices $H_1$ and $H_2$ of the same size, so $r_1=r_2=r$ and $n_1=n_2=n=2r$ (we make a choice to use matrices with half as many rows as columns).

\subsection{Types of Error Recovery Failures for the Pruned Peeling + VH Decoder}
\label{app:HGP_decoder}
In Sec.~\ref{sec_HGP_decoder} and Sec.~\ref{app:peeling_HGP}, we observed that the combined decoder (peeling + pruned peeling + VH) is not maximum-likelihood since decoder failures are possible in addtion to logical errors. Failure rate in the literature usually refers to the logical error rate, which is the only source of errors for a maximum-likelihood decoder. Logical errors for the erasure channel can only occur when the erasure covers a logical code word. However, there may exist erasure patterns covering a logical error which result in a decoder failure, and hence are not properly identified as logical errors. This distinction is stated visually by the Venn diagram of Fig.~\ref{fig:DF_LE}. The failure rate computed in our numerical simulations for the \textit{combined decoder} is the cummulative effect of these two possibilities, what we refer to as \textit{error recovery failure rate} on the vertical axis in the plots of our numerical simulations for HGP codes. Note that failures at low erasure rates are almost exclusively due to logical errors, and so this distinction can be regarded as negligible in the practical regime.

Note that peeling + pruned peeling is theoretically a maximum likelihood decoder in the special case of the surface code. This is equivalent to the spanning-tree-based ML decoder for the surface code~\cite{delfosse2020linear}. However, our implementation of pruned peeling is not perfect since it cannot identify the support of an arbitrary erased stabilizer. For the \textit{combined decoder}, the simplest classical stopping sets correspond to a fully erased row or column in the HGP Tanner graph. These are exactly the stopping sets of a repetition code, coinciding with logical errors for the surface code. In general, there do not exist erasure patterns giving a VH decoder failure for the surface code which do not also cover a logical error.

Figure \ref{fig:Toric10_VH_performance} shows the performance of the \textit{combined decoder} applied to the $10\times 10$ surface code. Comparing this to Fig.~\ref{fig:m}, which uses the ML decoder for the same surface code, we see a noticeable degradation in performance. This gap is explained by the existence of decoder failures in the combined case which do not exist for the ML decoder. Furthermore, the failure rate of the combined decoder converges to 1 as the erasure rate goes to 1, in contrast with the convergence to 0.75 for the ML decoder. This is because the erasure pattern is always a VH decoder stopping set when all qubits are erased, guaranteeing a decoder failure. Since there are no stopping sets in the ML case, however, a  100\% erasure rate is equivalent to generating a uniformly random physical Pauli error on the code. We see a convergence to 0.75 logical error rate because this error is identity 25\% of the time.

\begin{figure}[ht]
  \centering
  \begin{tikzpicture}
    \begin{scope} \clip (0,0) circle [radius=1];
      \fill[white!] (1,0) circle [radius=1];
      \fill[black!10!] (0,0) circle [radius=1];
    \end{scope}
    \draw (0,0) circle [radius=1];
    \draw (1,0) circle [radius=1];
    \draw ({cos(pi*3/4 r)},{sin(pi*3/4 r)}) node[fill=white]{DF};
    \draw ({1+cos(pi/4 r)},{sin(pi/4 r)}) node[fill=white]{LE};
    \draw [decorate,decoration={brace,amplitude=5pt,mirror,raise=6ex}]
  (-1,0) -- (2,0) node[midway,yshift=-4em]{ERF};
    
  \end{tikzpicture}
  \caption{Venn diagram distinguishing between the types of failures possible using the pruned peeling + VH decoder. A \textit{decoding failure} (DF) occurs when the decoder becomes stuck in a stopping set it cannot correct, in which case the algorithm fails to terminate. A \textit{logical error} (LE) occurs when the set of erased qubits covers a logical operator. An \textit{error recovery failure} (ERF) refers to the union of these two possibilities.
  It is possible that an application of the decoding algorithm to an erasure pattern covering a logical operator successfully returns to the code space, but results in a logical error; this possibility is represented in the Venn diagram by LE$\setminus$DF. Alternatively, DF$\cap$LE denotes those stopping sets for the decoder that cover a logical operator while DF$\setminus$LE denotes those stopping sets that do not. In all three cases, the decoder fails to recover the error.}
  \label{fig:DF_LE}
\end{figure}

\subsection{Technical Details for Assignment Strategies used with Multiplexed HGP Codes}
\label{app:HGP_strategies}

{\bf Strategy ii. stabilizer}
The stabilizer strategy was initially introduced in Sec.~\ref{sec_SC} for the surface code, but it can be applied to CSS codes more generally. The erased qubit-support of a stabilizer will be a peeling decoder stopping set, but these are precisely the stopping sets that the pruned peeling algorithm attempts to correct. Hence, this strategy is motivated by the idea that losing a photon corresponding to a single stabilizer individually induces a correctable erasure pattern.

In the stabilizer strategy for HGP codes, we partition the qubits into sets corresponding to the qubit-support for disjoint stabilizers. Qubits are assigned to stabilizers based on these sets. The number of qubits per stabilizer is fixed for an LDPC code and matches the row weight of $H_X$ or $H_Z$. Depending on the multiplexing number, the photons can also represent partial stabilizers or multiple stabilizers.

In the special case of the surface code with a $d\times d$ lattice, where $d$ is divisible by 4, it is always possible to partition the qubits into a combination of disjoint $X$- and $Z$-type stabilizer generators as seen in Fig.~\ref{fig:surface_stabilizer}, each of which is supported on 4 qubits. For a more general HGP code, we may attempt a similar assignment strategy by identifying the qubit-support of the stabilizer generators from the rows of $H_X$ and $H_Z$. However, we cannot guarantee that a partition of qubits into disjoint stabilizers is possible without placing constraints on the number of qubits and the row and column weights in the parity check matrices. Instead, we adopt an imperfect but simpler strategy for generic HGP codes, which does not require any additional assumptions about the code except that $H_X$ and $H_Z$ are LDPC. This strategy can be used with stabilizers coming only from $H_X$, only from $H_Z$, or a combination of both, provided that these matrices have the same row weight. Note that restricting to a single type of stabilizer creates a bias in the error correction, as was commented in the surface code case.

The first step of this strategy is to search for a partition of the qubits into disjoint stabilizers. To do this, we begin by choosing a row at random from $H_X$ or $H_Z$; the nonzero entries in this row represent the qubit-support of a single stabilizer. We then eliminate any overlapping stabilizers by deleting the rows from the matrices that share columns with nonzero entries with the previously selected row. Then we repeat this strategy until either all qubits have been divided into disjoint stabilizers or we exhaust the remaining rows that do not overlap with our previous selections. The result is that as many qubits as possible have been divided into non-overlapping sets corresponding to the qubit-support of disjoint stabilizers, possibly with some remaining ungrouped qubits.

Finally, the qubits are assigned to photons based on the disjoint sets identified in the previous step. Ordering the qubits by their stabilizer assignments, we then redistribute these into photons. The remaining ungrouped qubits are assigned after exhausting the chosen stabilizers. When the multiplexing number matches the stabilizer weight (that is, the row weight of $H_X$ or $H_Z$), each photon ideally matches a stabilizer, possibly with some remainder photons at the end for the ungrouped qubits. When the multiplexing number matches a fraction or multiple of the stabilizer weight, then the photons represent a partial stabilizer or multiple stabilizers, respectively. Allowing for the leftover qubits at the end ensures that this strategy can be applied with various multiplexing numbers, even when a perfect partition of qubits into stabilizers is not found. Because stabilizers are selected at random, this assignment strategy can be understood as a combination of the random and stabilizer strategies introduced before.

{\bf Strategy iii. sudoku}
The VH decoder is designed to address classical stopping sets for the peeling decoder, but there exist combinations of classical stopping sets that cannot be solved using this technique and result in a decoder failure. However, we may reduce the likelihood of a decoder failure by reducing the number of classical stopping sets in general. Classical stopping sets are supported on a single row or column of the HGP code Tanner graph. Thus, we propose an assignment strategy based on choosing qubits in the same photon from different rows and columns. The method for doing this is outlined in Algorithm~\ref{alg:sudoku}.

\begin{algorithm}[t]
    \caption{Strategy iii. sudoku}
    \label{alg:sudoku}
    \KwData{$P=\{p_i\}$ (the set of photons, where $p_i$ is the set of qubits in photon $i$), $Q=\{q_j=(r_j,c_j,b_j)\}$ (a list of 3-tuples with the row, column, and block of each physical qubit in the HGP code), and the number $m$ of qubits per photon.}
    \KwResult{$P=\{p_i\}$ (photon assignments).}
    
    \For{photon $p_i\in P$}{
        Pick a qubit $q_j\in Q$ at random\;
        Move $q_j$ from $Q$ to $p_i$\;
        \While{$|p_i|<m$ and $Q\neq\emptyset$}{
            Pick a candidate qubit $q_k\in Q$\;
            \eIf{$q_k$ is in a different row and column (or block) from each previously selected $q_j\in p_i$ (($r_k\neq r_j$ and $c_k\neq c_j$) or $b_k\neq b_j$)}{
                Move $q_k$ from $Q$ to $p_i$\;
            }{
                Move $q_k$ from $Q$ to a temporary waiting list $Q'$\;
            }   
        }
        Move all qubits in $Q'$ back to $Q$\;        \While{$|p_i|<m$}{
            Pick a qubit $q_k\in Q$ at random\;
            Move $q_k$ from $Q$ to $p_i$\;
        }
    }
    Return $P$;
\end{algorithm}

This strategy assumes that the number of qubits per photon does not exceed the minimum length of a row or column in the Tanner graph, although this condition may be relaxed by instead allowing for a minimal number of qubits from the same row or column to be added to the same photon. Qubits are assigned to photons at random, checking that each newly added qubit is not supported on the same row or column as any qubit already assigned to a given photon. In the case of a fixed number of photons where no valid qubit assignments remain, we drop the condition and default to random assignment.

{\bf Strategy iv: row-col}
Although not a practical assignment strategy, the case where only qubits from the same row or column of the HGP code Tanner graph are assigned to the same photon is of theoretical interest. This strategy attempts to maximize the number of classical stopping sets resulting from photon loss and thus increase the likelihood of a VH decoder failure. Verifying that this assignment strategy performs very poorly in numerical simulations serves as a proof of concept for the VH decoder and also justifies the preferred strategies using qubits from different rows and columns.

\begin{figure}
    \centering
    \includegraphics[width = 1 \columnwidth]{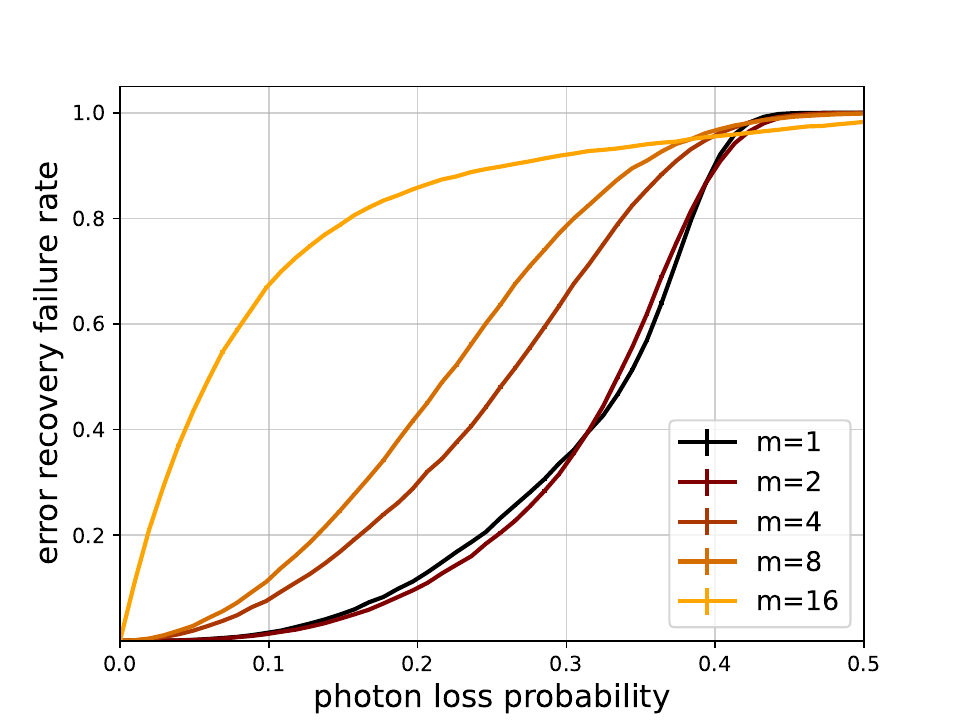}
    \caption{The performance for multiplexed communication with a [\![512,8]\!] equal-block ($16\times 16$) HGP code obtained from the symmetric construction using $r=n=16$ with assignment strategy (iv) row-column. Increasing the number of qubits per photon using this strategy rapidly increases the failure rate. At $m=16$, each photon corresponds to an entire row or column in the HGP Tanner graph, whereby losing even one photon guarantees a classical stopping set.}
    \label{fig:n512_as4}
\end{figure}

Fig.~\ref{fig:n512_as4} shows the performance of this strategy for a [\![512,8]\!] HGP code at several multiplexing numbers. Although surprisingly the $m=2$ case seems to outperform the no-multiplexing case, the failure rate otherwise increases as $m$ increases. Failures of the VH decoder are the result of certain configurations of classical stopping sets, and hence increasing the latter also increases the former. In particular, this explains the dramatic jump between the $m=8$ and $m=16$ cases. Since the blocks in this code's Tanner graph are $16\times 16$, each photon in the $m=16$ case corresponds to an entire row or column. Loss of any photon yields a classical stopping set, and hence VH decoder failures are common. This also confirms the significance of designing assignment strategies to avoid classical stopping sets in our simulations of HGP codes. In general, we expect the performance of the row-column strategy to drop significantly as $m$ becomes equal to or larger than the side length of the block in the HGP Tanner graph.

{\bf Strategy v. diagonal}
Whereas the sudoku strategy assigns qubits at random subject to the condition of being in a different row or column, qubits in the HGP code Tanner graph may also be grouped diagonally within each block. In this way, it is possible to satisfy the sudoku condition without relying on randomness. A $d\times d$ grid can be divided into $d$ non-overlapping diagonal slices, where we allow slices to wrap around. Since no two qubits in the same diagonal slice are contained in the same row or column of the grid, this technique also guarantees that we avoid classical stopping sets within a single photon. Photon assignment is thus based on grouping together the qubits in the same diagonal slice. Each of the two qubit-squares in the HGP code Tanner graph is considered separately, but if we require that the ratio of the squares' side lengths is a whole number, then the qubits can be cleanly partitioned into photons of size matching the side length of the smaller square. HGP codes with rectangular Tanner graph block sizes can also use the diagonal strategy, provided that the length of the diagonal slice does not exceed the length of the shortest side. If longer slices are permitted in the rectangular case, then instead a minimal number of qubits in the same row or column are allowed.

The implementation of this strategy as described in Algorithm~\ref{alg:diagonal} is simple, provided one precomputes a \textit{diagonal ordering} on the qubits in the HGP Tanner graph. Referring to the block structure of Fig.~\ref{fig:HGP_TannerGraph}, the qubits in a given block are indexed along the non-overlapping diagonal slices. These slices are allowed to wrap around the sides of the square, which guarantees that no two qubits in the same slice are contained in the same row or column. The qubits in the second block are indexed sequentially after the first block. In our numerical implementation, a separate function to compute this ordering on the qubits in an HGP code is used along with the assignment function.

\begin{algorithm}
    \caption{Strategy v. diagonal}
    \label{alg:diagonal}
    \KwData{$P=\{p_i\}$ (the set of photons where $p_i$ is the set of qubits in photon $i$), $Q=\{q_j\}$ (a list of physical qubits ordered along the diagonal), and the number $m$ of qubits per photon.}
    \KwResult{$P=\{p_i\}$ (photon assignments).}
    \For{photon $p_i\in P$}{
        \For{qubits with indices $j\in\{im,\cdots,(i+1)m\}$}{
            Move $q_j$ from $Q$ to $p_i$
        }
    }
    return $P$\;
\end{algorithm}

\end{document}